\documentclass[11pt,a4paper]{article}
\usepackage[utf8]{inputenc}
\usepackage{amsmath}
\usepackage{amsfonts}
\usepackage{amssymb}
\usepackage{makeidx}
\usepackage{graphicx}
\usepackage{mathabx} 
\usepackage{dsfont} 
\usepackage[left=2cm,right=2cm,top=2cm,bottom=2cm]{geometry}
\usepackage{color}
\usepackage{verbatim} 
\usepackage[nottoc]{tocbibind} 
\usepackage{hyperref} 
\usepackage{mathtools} 
\usepackage{amsthm}
\usepackage{cancel}
\usepackage{float}
\usepackage{appendix}
\usepackage{emptypage} 

\numberwithin{equation}{section}

\newtheorem{theorem}{Theorem}[section]
\newtheorem{definition}{Definition}[section]

\begin{document}

\setlength{\parindent}{0pt} 

\begin{center}
\Large\textbf{MODELING OF FLUID-RIGID BODY INTERACTION IN AN ELECTRICALLY CONDUCTING FLUID} \\[10mm]
\large{JAN SCHERZ}$^{1,2,3}$,\ \large{ANJA SCHLÖMERKEMPER}$^1$ \\[10mm]
\end{center}

\begin{itemize}
\item[$^1$] Institute of Mathematics, University of Würzburg, Emil-Fischer-Str. 40, 97074 Würzburg, Germany
\item[$^2$] Department of Mathematical Analysis, Faculty of Mathematics and Physics, Charles University in Prague, Sokolovská 83, Prague 8, 18675, Czech Republic
\item[$^3$] Mathematical Institute, Czech Academy of Sciences, \v{Z}itná 25, Prague 1, 11567, Czech Republic
\end{itemize} 

\begin{center}
\Large\textbf{Abstract} \\[4mm]
\end{center}
      
We derive a mathematical model for the motion of several insulating rigid bodies through an electrically conducting fluid. Starting from a universal model describing this phenomenon in generality, we elaborate (simplifying) physical assumptions under which a mathematical analysis of the model becomes feasible. Our main focus lies on the derivation of the boundary and interface conditions for the electromagnetic fields as well as the derivation of the magnetohydrodynamic approximation carried out via a nondimensionalization of the system.

\section{Introduction}

The study of electrically conducting fluids interacting with solid materials is motivated by biomechanical applications such as capsule endoscopy, cf.\ \cite{endocapsules}. This medical procedure constitutes a minimally invasive method for the detection of diseases by propelling small capsule shaped camera devices through parts of the human body such as veins or arteries. In the electrically conducting blood it is possible to generate the drive and control the navigation of these devices remotely by applying electromagnetic forces. A similar application is given by a procedure known as remote drug delivery, cf.\ \cite[Section 4.4]{drugdelivery}: Robots of a microscopic scale can be used to transport medication through the blood stream in the human body directly to the region of the body in which it is needed while damage through the medication to healthy tissue is avoided.

In this article we study the mathematical modeling
of the motion of several insulating
rigid bodies through an electrically conducting diamagnetic dielectric viscous non-homogeneous fluid
surrounded by a perfect conductor. In the mathematical examination of this and similar phenomena, the considered models are typically simplified subject to certain physical restrictions in order to make their mathematical analysis feasible. The main contribution of the present article is the rigorous derivation of specific models used in the literature (see below) from a universal model describing the situation in generality. Further, the physical assumptions under which this derivation is justified are elaborated on.

The interaction between electrically conducting fluids and insulating rigid bodies constitutes a combination of two types of physical phenomena; fluid-structure interaction (FSI) and magnetohydrodynamics (MHD). The research field of FSI deals with interactions between rigid or deformable solids and fluids contained in, adjacent to or surrounding the solids, cf.\ \cite{fsiintro,bungartzschafer}. In the case of rigid bodies, FSI is commonly modeled via a coupling of the (incompressible or compressible) Navier-Stokes equations for the description of the fluid motion and the balances of linear and angular momentum of the bodies. The mathematical literature is rich in analytical results on this model: We specifically mention the articles \cite{feireisl,incompressiblefeireisl}, wherein the global-in-time existence of weak solutions to the problem was proved in the compressible and the incompressible setting, respectively. We further mention the article \cite{tucsnak}, in which the same result was shown in the two-dimensional incompressible setting with an additional examination of the question of the possibility of collisions between the bodies. 

MHD instead describes the study of the interaction between electrically conducting fluids and electromagnetic fields. The behaviour of electromagnetic fields is described by the Maxwell equations, cf.\ for example \cite{eringenmaugin}. In MHD, these equations are then coupled with the Navier-Stokes equations for the description of the fluid, cf.\ \cite{cabannes,davidson,kl}. The equations in this coupling are further simplified in comparison to the original systems under several physical assumptions. The latter procedure is commonly referred to as the magnetohydrodynamic approximation. A justification of this approximation is for example given in \cite{jiliang,jiliang2}. Mathematically, the MHD problem is also well investigated and we refer to \cite{gbl,sart} for the proofs of the global-in-time existence of weak solutions in the case of an incompressible and a compressible fluid, respectively. 

Despite the extensively worked out theory in both FSI and MHD, not many results on the combination between these two research areas appear to be available. A first mathematical exploration of this uncharted territory was made in \cite{guermondminev2d,guermondminev}. In those articles the flow of an electrically conducting incompressible fluid around an insulating rigid body is studied in two and three spatial dimensions, respectively, and the existence of weak solutions to the corresponding models is proved. The rigid body in those articles, however, is assumed to be immovable; an extension of the result to the case of a freely moving body was achieved by the authors of the present article, in joint work with Barbora Bene\v{s}ová and \v{S}árka Ne\v{c}asová, in \cite{incompressiblecase}. A further extension to the setting of a compressible fluid was obtained later in \cite{compressiblepaper}, see also \cite{thesis}.

In the present article we derive the models used in the latter two papers (and thus of \cite{guermondminev2d,guermondminev} when assuming that the rigid bodies are not moveable anymore). More precisely, we provide the mathematical derivation of a model for  the motion of several insulating rigid bodies through an electrically conducting diamagnetic dielectric viscous non-homogeneous fluid surrounded by a perfect conductor. The individual partial differential equations in this model, e.g. the Navier-Stokes equations and the Maxwell's equations, are well-known. In particular, as mentioned above, there exists plenty of literature on magnetohydrodynamics (see again \cite{cabannes,davidson,kl}), which lies in the center of interest of our investigations. We thus remark that the findings of this article do not constitute a new result. Instead, we provide deeper mathematical insight in the derivation than we were able to find in the literature.

We further point out that the present article is a self-contained variant of a chapter from the PhD thesis \cite{thesis} of the first author, in which the derivation of the desired model is given in the case of an incompressible fluid. Here, we focus on the compressible case, which is achieved by the same procedure as in the incompressible case.

The outline of the article is as follows: In Section \ref{finalmodel}, we present the desired model analyzed mathematically in \cite{compressiblepaper}, the mathematical derivation of which is the main contribution of this article. Our starting point in the derivation of this model is a more general system composed of the Maxwell equations, the compressible Navier-Stokes equations and the balances of linear and angular momentum of the rigid bodies in their most universal forms, which we present in Section \ref{originalmodel}. This system models the interaction between the fluid, the rigid bodies and the electromagnetic fields inside of both materials in full generality. After a first adjustment of the system to the properties of the materials we consider in our specific setting - in particular the non-conductivity of the solid domain - our procedure consists of two main steps: Firstly, in Section \ref{bicond}, we derive boundary and interface conditions for the electromagnetic fields from the Maxwell equations. Secondly, in Section \ref{non-dimensionalization}, we simplify the system by identifying several insignificant terms and dropping them from the equations. This step, which is achieved via a nondimensionalization, in particular constitutes the classical magnetohydrodynamic approximation in the fluid part of the domain. In Section \ref{finalsystem}, we summarize the final system we achieve through this approach. This system constitutes a slightly more general version of the model from Section \ref{finalmodel}. The latter system is then obtained after a few more (mathematical) adjustments discussed at the end of Section \ref{finalsystem}. Finally, in Section \ref{exresult}, we give a short recap of the result 
on the existence of weak solutions to the derived model, which was originally stated in \cite{compressiblepaper}.

\section{Model used in the mathematical analysis} \label{finalmodel}

We study several insulating rigid bodies moving through an electrically conducting viscous non-homogeneous compressible fluid. Additionally, we assume the fluid to be a linearly magnetic material and a linearly dielectric, cf.\ the relations \eqref{304} below. The latter conditions are in particular satisfied for diamagnetic linearly dielectrics such as for example blood, cf.\ \cite{diamagnetic,dielectric}. We also assume the fluid to be surrounded by a (rigid) perfect conductor. In the present section we showcase a mathematical model of this situation which has been analyzed mathematically in \cite{compressiblepaper}. This model, however, is subject to certain further physical and mathematical assumptions. The main contribution of this article is the derivation of this model from a universal model describing the above situation in full generality, cf.\ Section \ref{originalmodel} below.

We consider a domain $\Omega \subset \mathbb{R}^3$, a time interval $[0,T]$ with $T>0$ and we denote by $Q$ the time space domain $Q := (0,T)\times \Omega$. The domain $\Omega$ is filled, at each time $t \in [0,T]$, with $N \in \mathbb{N}$ rigid bodies $S^i(t) \subset \Omega$, $i=1,...,N$ as well as the fluid $F(t) := \Omega \setminus \overline{S(t)}$, where
\begin{align}
S(t) := \bigcup_{i=1}^N S^i(t) \nonumber
\end{align}

denotes the solid domain at time $t$. Similarly, we divide the time space domain $Q$ into its solid part $Q^s$ and its fluid part $Q^f$,
\begin{align}
Q^{s} := \left\lbrace (t,x) \in Q :\ x \in S(t) \right\rbrace,\quad \quad Q^f := Q \setminus \overline{Q^s}. \nonumber
\end{align}

For any function defined on $Q$ we indicate, whenever it is necessary to stress the difference, its restrictions to $Q^s$ and $Q^f$ by the superscripts $s$ and $f$, respectively. The quantities which model the interaction between the electromagnetic fields, the fluid and the rigid bodies can be distinguished between electromagnetic and mechanical quantities. On the electromagnetic side we have the magnetic induction $B:Q \rightarrow \mathbb{R}^3$, the magnetic field $H:Q \rightarrow \mathbb{R}^3$, the electric field $E:Q \rightarrow \mathbb{R}^3$ and the electric current density $j:Q \rightarrow \mathbb{R}^3$. The mechanical quantities are the density $\rho:Q \rightarrow \mathbb{R}$, the velocity field $u:Q \rightarrow \mathbb{R}^3$ and the pressure $p = p^f:Q^f \rightarrow \mathbb{R}$. The evolution of these quantities is determined by a system of partial differential equations consisting of the Maxwell equations
\begin{align}
\text{curl} H &= \left\{
                \begin{array}{ll}
                  j + J \ \ \ &\text{in } Q^f,\\
                  0 &\text{in } Q^s,
                \end{array}
              \right. \label{2919} \\
\partial_t B + \text{curl} E &= 0 \quad \quad \quad \quad \quad \text{in } Q^f \ \text{and } Q^s, \label{2904} \\\
\text{div} E &= 0 \quad \quad \quad \quad \quad \text{in } Q^s, \label{2918} \\
\text{div} B &= 0 \quad \quad \quad \quad \quad \text{in } Q^f \ \text{and } Q^s, \label{2917}
\end{align}

the compressible Navier-Stokes equations
\begin{align}
\partial _t \rho + \operatorname{div} (\rho u) =& 0 \quad \quad \quad \quad \quad \quad \quad \quad \quad \quad \quad \quad \ \text{in } Q^f, \label{2916} \\
\partial _{t} (\rho u) + \text{div} (\rho u \otimes u) + \nabla p =& \text{div} \mathbb{T} + \rho g + \frac{1}{\mu} \text{curl}B\times B\quad \quad \text{in } Q^f \label{2915}
\end{align}

and the balances of linear and angular momentum
\begin{align}
m^i \frac{d}{dt} V^i(t) = \frac{d}{dt} \int_{S^i(t)} \rho u\ dx = \int_{\partial S^i(t)} \left[\mathbb{T} - p\ \text{id}\right] \text{n}\ dA + \int _{S^i(t)} \rho g \ dx,& \quad t \in [0,T], \label{2914} \\
\frac{d}{dt}\left( \mathbb{J}^i(t)w^i(t) \right) = \frac{d}{dt} \int _{S^i(t)} \rho \left(x - X^i\right) \times u \ dx\ & \nonumber \\
= \int_{\partial S^i(t)} (x - X^i) \times \left[\mathbb{T} - p\ \text{id} \right] \text{n}\ dA + \int _{S^i(t)} \rho \left(x - X^i\right) \times g\ dx,& \quad t \in [0,T] \label{2913}
\end{align}

for $i=1,...,N$ in combination with the relations
\begin{align}
&j = \sigma (E + u \times B) \quad \text{in } Q^f \ \text{and } Q^s,\quad \quad \sigma = \left\{
                \begin{array}{ll}
                  \sigma ^f > 0 \ \ \ &\text{in } Q^f,\\
                  \sigma ^s = 0 &\text{in } Q^s,
                \end{array}
              \right. \label{2912} \\
&\quad \quad \quad \quad \quad \quad \ \ B = \mu H, \quad \quad \mu > 0 \quad \text{in } Q \label{2911}
\end{align}

as well as the boundary and interface conditions
\begin{align}
B(t) \cdot \text{n} &= 0\quad \text{on } \partial \Omega,\quad \quad &&\ \ \ \ \ \ \ \ \ B^f(t) - B^s(t) = 0 \quad \text{on } \partial S(t), \label{2910} \\
E(t) \times \text{n} &= 0 \quad \text{on } \partial \Omega,\quad \quad &&\left(E^f(t) - E^s(t)\right) \times \text{n} = 0 \quad \text{on } \partial S(t), \label{2909} \\
u(t) &= 0\quad \text{on } \partial \Omega,\quad \quad &&\ \ \ \ \ \ \ \ \ \ u^f(t) - u^s(t) = 0 \quad \text{on } \partial S(t). \label{2908}
\end{align}

The Maxwell system \eqref{2919}--\eqref{2917} consists of Ampère's law \eqref{2919} - in which $J:Q \rightarrow \mathbb{R}^3$ constitutes an external forcing term -, the Maxwell-Faraday equation \eqref{2904}, Gauss's law \eqref{2918} and Gauss's law for magnetism \eqref{2917}. The Navier-Stokes system \eqref{2916}, \eqref{2915} is composed of the continuity equation \eqref{2916} and the momentum equation \eqref{2915}. The latter equation contains two forcing terms: On the one hand, the quantity $\rho g$, in which the $g:Q \rightarrow \mathbb{R}^3$ denotes a given function, represents an external force, e.g.\ the gravitational force. On the other hand, the influence of the electromagnetic fields on the motion of the fluid shows itself in the Lorentz force $\frac{1}{\mu} \operatorname{curl} B \times B$. The stress tensor $\mathbb{T}$ in the momentum equation is defined as
\begin{align}
\mathbb{T} = \mathbb{T}(u) := 2\nu \mathbb{D}(u) + \lambda \text{id}\ \operatorname{div} u,\quad \quad \mathbb{D}(u) := \frac{1}{2} \nabla u + \frac{1}{2} (\nabla u)^T, \nonumber
\end{align}

where the constants $\nu > 0$ and $\lambda \geq - \nu$ are referred to as the viscosity coefficients. Moreover, the pressure $p$ in the momentum equation is assumed to depend only on the density $\rho$. More specifically, it is assumed to satisfy the isentropic constitutive relation
\begin{align}
p = p \left(\rho^f \right) = a \left( \rho^f \right)^\gamma, \quad a>0,\ \gamma > \frac{3}{2}. \label{2907}
\end{align}

In the balances \eqref{2914}, \eqref{2913} of linear and angular momentum, the quantities
\begin{align}
m ^i:=& \int _{S^i(t)} \rho (t,x)\ dx,\quad X^i(t):= \frac{1}{m^i}\int _{S^i(t)} \rho (t,x)x\ dx, \nonumber \\
\mathbb{J}^i(t)a \cdot b :=& \int _{S^i(t)}\rho (t,x)\left[ a \times \left( x - X^i(t) \right) \right] \cdot \left[ b \times \left( x - X^i(t) \right) \right]\ dx,\quad a,b \in \mathbb{R}^3, \nonumber
\end{align}

denote the mass $m^i$, the center of mass $X^i$ and the inertia tensor $\mathbb{J}^i$ of the $i$-th rigid body, respectively. These equations determine the translational velocity $V^i$ as well as the rotational velocity $w^i$ of the $i$-th rigid body, the overall velocity of which is then given as
\begin{align}
u(t,x) = u^{s^i}(t,x) := V^i(t) + w^i(t) \times \left( x - X^i(t) \right) \quad \quad \text{for all } t \in [0,T]\ \text{and } x \in S^i(t). \nonumber
\end{align}

In particular, the balances of linear and angular momentum show how the motion of the insulating bodies is influenced by the fluid, however, not - directly - by the electromagnetic fields. The influence of the fluid on the electromagnetic fields is seen in Ohm's law \eqref{2912}. In this relation, $\sigma$ stands for the electric conductivity. Since $\sigma = \sigma^s = 0$ in the insulating solids, the equation further shows that the bodies have no (direct) impact on the behavior of the electromagnetic fields. The constant $\mu$ in the linear relation \eqref{2911} represents the magnetic permeability and, finally, we point out that in the interface and boundary conditions \eqref{2910}--\eqref{2908} the no slip condition \eqref{2908} for the velocity field displays the influence of the rigid bodies on the fluid motion.

As explained above, the system \eqref{2919}--\eqref{2908} does not model the interaction between insulating rigid bodies and an electrically conducting fluid in full generality but rather under certain additional assumptions. In the following section we will present a universal model and subsequently we will precisely work out the assumptions under which this model can be modified such that it turns into the model \eqref{2919}--\eqref{2908}.

\section{General model} \label{originalmodel}

In the present section we showcase a universal model for the motion of insulating rigid bodies through an electrically conducting linearly dielectric and linearly magnetic viscous non-homogeneous compressible fluid surrounded by a perfect conductor. The notation for this model essentially coincides with the one for the model \eqref{2919}--\eqref{2908}, however, it requires a few additions which we summarize in the following. The domain of the perfect conductor is $\mathbb{R}^3 \setminus \overline{\Omega}$, we denote the associated time space domain by
\begin{align}
Q^{\operatorname{ext}} := (0,T) \times \left( \mathbb{R}^3 \setminus \overline{\Omega} \right). \nonumber
\end{align}

The restriction of functions defined on $(0,T)\times \mathbb{R}^3$ to $Q^{\operatorname{ext}}$ is indicated by the superscript $\operatorname{ext}$. Further, for the desription of the electromagnetic effects in the general setting, we require several additional physical quantities: We denote by $D:(0,T)\times \mathbb{R}^3\rightarrow \mathbb{R}^3$ the electric induction, by $\rho_c:(0,T)\times \mathbb{R}^3\rightarrow \mathbb{R}$ the density of electric charges, by $M:(0,T)\times \mathbb{R}^3\rightarrow \mathbb{R}^3$ the magnetization and by $P:(0,T)\times \mathbb{R}^3\rightarrow \mathbb{R}^3$ the polarization. The system of partial differential equations modelling the considered fluid-rigid body interaction in full generality consists of the Maxwell equations
\begin{align}
\operatorname{curl} H  =& \partial _t D  + j + J \quad \quad \text{in } Q^f,\ Q^s \text{ and } Q^{\text{ext}}, \label{293} \\
\partial _t B + \operatorname{curl} E =& 0 \quad \quad \quad \quad \quad \quad \ \ \text{in } Q^f,\ Q^s \text{ and } Q^{\text{ext}}, \label{292} \\
\operatorname{div} D =& \rho _c \quad \quad \quad \quad \quad \quad \ \text{in } Q^f,\ Q^s \text{ and } Q^{\text{ext}}, \label{290} \\
\operatorname{div} B =& 0 \quad \quad \quad \quad \quad \quad \ \ \text{in } Q^f,\ Q^s \text{ and }Q^{\text{ext}}, \label{291}
\end{align}

the compressible Navier-Stokes equations
\begin{align}
\partial _t \rho + \operatorname{div} \left( \rho u \right) =& 0 \quad \quad \quad \quad \quad \quad \quad \quad \quad \quad \quad \quad \quad \quad \quad \ \text{in } Q^f, \label{1.1} \\
\partial_t \left(\rho u\right) + \operatorname{div} \left(\rho u \otimes u\right) + \nabla p =& \operatorname{div} \mathbb{T} + \rho g + \rho _c E + \left(j + J \right) \times B\quad \quad \text{in } Q^f, \label{1.2}
\end{align}

and the balances of linear and angular momentum
\begin{align}
m^i \frac{d}{dt} V^i(t) = \frac{d}{dt} \int_{S^i(t)} \rho u\ dx = \int_{\partial S^i(t)} \left[\mathbb{T} - p\ \text{id}\right] \text{n}\ dA + \int _{S^i(t)} \rho g \ dx,& \quad t \in [0,T], \label{1.7} \\
\frac{d}{dt}\left( \mathbb{J}^i(t)w^i(t) \right) = \frac{d}{dt} \int _{S^i(t)} \rho \left(x - X^i\right) \times u \ dx\ & \nonumber \\
= \int_{\partial S^i(t)} (x - X^i) \times \left[\mathbb{T} - p\ \text{id} \right] \text{n}\ dA + \int _{S^i(t)} \rho \left(x - X^i\right) \times g\ dx,& \quad t \in [0,T] \label{1.9}
\end{align}

for $i=1,...,N$ in combination with the relations
\begin{align}
j = \left\{
                \begin{array}{ll}
                 \sigma (E + u \times B) \ \ \ &\text{in } Q^f\ \text{and } Q^s,\\
                 \sigma E &\text{in } Q^{\text{ext}},
                \end{array}
              \right.& \quad \quad \sigma = \left\{
                \begin{array}{ll}
                  \ \ \sigma ^f > 0 \ \ \ &\text{in } Q^f,\\
                  \ \ \sigma ^s := 0 &\text{in } Q^s,\\
                  \sigma^{\text{ext}} := +\infty & \text{in } Q^{\text{ext}},
                \end{array}
              \right. \label{306} \\
H = \frac{1}{\mu_0} B - M \quad \quad \text{in } Q^f,\ Q^s \text{ and } Q^{\text{ext}}, \quad \quad \quad& \quad D= \epsilon_0 E + P \quad \quad \text{in } Q^f,\ Q^s \text{ and } Q^{\text{ext}} \label{300}
\end{align}

as well as the boundary and interface conditions
\begin{align}
u(t) = 0\quad \quad \text{on } \partial \Omega,\quad \quad \quad \quad u^f(t) - u^s(t) = 0 \quad \quad &\text{on } \partial S(t). \label{noslip}
\end{align}

The most prominent differences in this system as opposed to the system \eqref{2919}--\eqref{2908} lie in the more general form of the Maxwell equations \eqref{293}--\eqref{291} and the momentum equation \eqref{1.2}. All of the Maxwell equations are now formulated in all three time space domains $Q^s$, $Q^f$ and $Q^{\operatorname{ext}}$ and they contain the additional quantities $\partial_t D$ and $\rho_c$ on the right-hand sides of Ampère's law \eqref{293} and Gauss's law \eqref{290}, respectively. In the momentum equation, the Lorentz force is now to be found in its general form $\rho_cE + (j + J) \times B$. Additionally, on the external force $J$ we here impose the assumptions
\begin{align}
\operatorname{div} J = 0 \quad \quad \text{in } Q^f,\quad \quad \quad \quad J = 0 \quad \quad \text{in } Q^s \text{ and } Q^{\text{ext}}. \label{assumptionsJ}
\end{align}

In Ohm's law \eqref{306} in the exterior domain, the choice $\sigma = \sigma^{\operatorname{ext}}=+\infty$ is in accordance with the exterior domain being occupied by a perfect conductor. The absence of the velocity field $u$ in this relation is explained by the fact that no motion occurs in this part of the domain. In the relations \eqref{300}, the first identity constitutes the general form of the linear relation \eqref{2911} between $B$ and $H$, in which $\mu_0$ denotes the value of the magnetic permeability $\mu$ in vaccum. The second identity in \eqref{300}, in which $\epsilon_0$ stands for the value of the dielectric permittivity in vacuum, constitutes a corresponding relation between $D$ and $E$. Finally, we point out that the boundary and interface conditions \eqref{2910}, \eqref{2909} for the electromagnetic fields are neglected in the general system. This is because these conditions, as opposed to the no-slip condition \eqref{noslip} for the velocity field, which constitutes a standard assumption in fluid-structure interaction models (see for example \cite{incompressiblefeireisl,tucsnak}), are inherent to the Maxwell equations. This can be seen in Section \ref{bicond} below, where we derive them explicitly.

\subsection{Mathematical assumptions}

For the mathematical calculations in the following sections we impose certain regularity assumptions on the involved functions and domains. These assumptions can be summarized as
\begin{align}
\text{The d}&\text{omains}\ \Omega \ \text{and}\ S(t)\ \text{are of class}\ C^{2,1}\ \text{at each time } t \in [0,T], \label{mathassumption1} \\
\rho, u, B,H,D,E&\ \text{and}\ J\ \text{are twice continuously differentiable in}\ Q^f,Q^s\ \text{and}\ Q^{\text{ext}}\ \text{and} \label{mathassumption2} \\
\rho&,u,B,H,D,E,J\ \text{and all their derivatives are bounded.} \label{mathassumption3}
\end{align}
We mainly require these assumptions for the derivation of the interface conditions of the electromagnetic fields in Section \ref{bicond} below. These conditions are derived from integrated versions of the Maxwell equations. The assumptions \eqref{mathassumption1}--\eqref{mathassumption3} are used for passing to the limit in the equations when the domain of integration shrinks to a point.

\subsection{Adjustments to the material assumptions} \label{materialsimplifications}

The physical properties assumed for the materials in consideration lead to some immediate simplifications of certain aspects of the model \eqref{293}--\eqref{noslip}. In the perfect conductor in the exterior domain we may assume, in accordance with the physical literature (cf.\ \cite[Chapter 1, Part A, §4.2.4.3]{dautraylions}), that
\begin{align}
B = E = j = 0\quad \text{in } Q^{\text{ext}}. \label{vanishingfields}
\end{align}
Indeed, in the (immovable) exterior domain we can set $u=0$. Moreover, by definition of the electrical conductivity $\sigma$ in \eqref{300}, we know that $\sigma = \sigma^{\text{ext}}=+\infty$ in $Q^{\text{ext}}$. Thus, due to the boundedness of $E$ assumed in \eqref{mathassumption3}, Ohm's law \eqref{306} implies that $E=j=0$. The Maxwell-Faraday equation \eqref{292} then further shows that $\partial_t B=0$. Assuming, without loss of generality, that $B=B(0)=0$ at the initial time $t=0$, we thus infer that $B=B(t)=0$ at all times $t \in [0,T]$. As a consequence of the trivial relations \eqref{vanishingfields} the equations for $B$, $E$ and $j$ in the exterior domain become superfluous in our system. In particular, under exploitation of the assumption \eqref{assumptionsJ} on $J$, the Maxwell system \eqref{293}--\eqref{291} in the exterior domain reduces to the equations
\begin{align}
\operatorname{curl} H = \partial_{t} D \quad \quad &\text{in } Q^{\text{ext}}, \label{ampext} \\
\operatorname{div} D = \rho _c \ \ \ \quad \quad &\text{in } Q^{\text{ext}}. \label{gaussext}
\end{align}
In the insulating solid domain the electrical conductivity $\sigma = \sigma^s$ vanishes, cf.\ \eqref{300}. Consequently, by Ohm's law \eqref{306}, it holds that $j = 0$ in $Q^s$. The non-conductivity of the solid also allows us to regard this region as a vacuum from the electromagnetic point of view, within which there exist no electric charges. In particular this means that the electric charge density vanishes in the solid region, i.e.\ it holds that
\begin{align}
\rho_c = 0 \quad \quad \text{in } Q^s. \label{4031}
\end{align}
Further, by the assumptions \eqref{assumptionsJ}, we know that $J = 0$ in $Q^s$. Under consideration of these facts, the Maxwell system \eqref{293}--\eqref{291} in the solid domain reduces to
\begin{align}
\operatorname{curl} H = \partial _t D \quad \quad &\text{in } Q^s, \label{1004} \\
\partial _t B + \operatorname{curl} E = 0 \quad \quad \quad &\text{in } Q^s, \label{1003} \\
\operatorname{div} D = 0 \quad \quad \quad &\text{in } Q^s, \label{1001} \\
\operatorname{div} B = 0 \quad \quad \quad &\text{in } Q^s. \label{1002}
\end{align}
The fact that the solid domain is considered as a vacuum from the electromagnetic point of view moreover means that the magnetization and the polarization vanish in this region,
\begin{align}
M = P = 0\quad \quad \text{in } Q^s. \label{311}
\end{align}
In the fluid domain, the assumption of the fluid being a linear (magnetic) material and a linear dielectric imply a linear relation between the magnetization and the magnetic field as well as between the polarization and the electric field, respectively. Namely, it holds that
\begin{align}
M = \chi_m^f H \quad \quad \text{in } Q^f,\quad \quad \quad \quad P = \epsilon_0 \chi_{e}^f E \quad \quad \text{in } Q^f, \label{312}
\end{align}
where $-1 \ll \chi_m^f <0$ and $0 < \chi_e^f$ denote the magnetic and the electric susceptibility of the fluid, respectively, cf.\ \cite[Section 4.4.1, Section 6.4.1]{griffiths}). The latter quantities are related to the magnetic permeability $\mu^f$ and the dielectric permittivity $\epsilon^f$ of the fluid via the relations
\begin{align}
\mu^f = \mu_0 \mu_r^f > 0,\quad \quad \epsilon^f = \epsilon_0 \epsilon_r^r > 0, \nonumber
\end{align}

where $\mu_r^f := 1 +\chi_m^f$ denotes the relative permeability and $\epsilon_r^f := 1 + \chi_e^f$ the relative permittivity of the fluid. Since the magnetic permeability and the dielectric permittivity in the insulating solid region coincide with the magnetic permeability $\mu_0$ and the dielectric permittivity $\epsilon_0$ in vacuum, respectively, we may thus write
\begin{align}
\mu = \left\{
                \begin{array}{ll}
                  \mu ^f = \mu_0 \mu_r^f \ \ \ &\text{in } Q^f,\\
                  \mu ^s = \mu_0 &\text{in } Q^s,
                \end{array}
              \right. \quad \quad \epsilon = \left\{
                \begin{array}{ll}
                  \epsilon ^f = \epsilon_0 \epsilon_r^f \ \ \ &\text{in } Q^f,\\
                  \epsilon ^s = \epsilon_0 &\text{in } Q^s.
                \end{array}
              \right. \label{muepsilon}
\end{align}

This, together with the relations \eqref{311} and \eqref{312}, reduces the nonlinear identities \eqref{300} in the fluid and the solid region to the linear equations
\begin{equation}
B = \mu H \quad \quad \text{in } Q^f \ \text{and}\ Q^s,\quad \quad \quad \quad D= \epsilon E\quad \quad \text{in } Q^f \ \text{and}\ Q^s. \label{304}
\end{equation}

\section{Interface conditions for the electromagnetic fields} \label{bicond}

\begin{figure}
\centering
\includegraphics[scale=2]{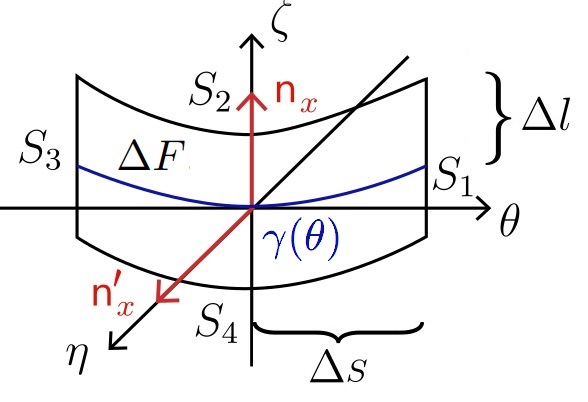} 
\hspace*{0.3cm}
\caption{The curved rectangle $\Delta F$.} \label{curvedrectangle}
\end{figure}

In this section we derive the interface conditions for the electromagnetic fields on $\partial \Omega$ and $\partial S(t)$, $t \in [0,T]$. We emphasize, that here we do not refer to the conditions on $\partial \Omega$ as boundary conditions. This is because we study the electromagnetic fields not only inside of $\Omega$ but also in the perfect conductor $\mathbb{R}^3 \setminus \overline{\Omega}$. The full set of interface conditions for the electromagnetic fields reads
\begin{align}
E^f(t) \times \text{n} &= 0 \quad \quad \ \text{on } \partial \Omega,\quad \quad &&\left(E^f(t) - E^s(t)\right) \times \text{n} = 0 \quad \quad \ \ \text{on } \partial S(t), \label{831mod} \\
\left(H^{\text{ext}}(t) - H^f(t) \right) \times \text{n} &= k(t) \quad \text{on } \partial \Omega,\quad \quad &&\left(H^f(t) - H^s(t)\right) \times \text{n} = 0 \quad \quad \ \text{on } \partial S(t), \label{832mod} \\
B^f(t) \cdot \text{n} &= 0\quad \quad \ \text{on } \partial \Omega,\quad \quad &&\ \ \left(B^f(t) - B^s(t)\right) \cdot \text{n} = 0 \quad \ \quad \text{on } \partial S(t), \label{833mod} \\
\left(D^{\text{ext}}(t) - D^f(t)\right) \cdot \text{n} &= \omega(t) \quad \text{on } \partial \Omega,\quad \quad &&\ \ \left(D^f(t) - D^s(t)\right) \cdot \text{n} = \omega(t) \quad \text{on } \partial S(t), \label{834mod}
\end{align}
for the surface current density $k$ on $\partial \Omega$ and the surface charge density $\omega$ on $\partial \Omega$ and $\partial S(t)$. We remark that here the surface charge density $\omega$ stands for the charge per unit area, whereas the surface current density $k$ is defined as the current per unit length. The latter definition becomes clear in the derivation of the condition \eqref{832mod} below, cf.\ the formula \eqref{surfacecurrentdensity}. We further remark that the fields $E^{\text{ext}}$ and $B^{\text{ext}}$ do not appear in the conditions on $\partial \Omega$ in \eqref{831mod} and \eqref{833mod}, which is due to the fact that these fields vanish in the exterior domain, cf.\ \eqref{vanishingfields}. In the following we deduce the above conditions from the Maxwell equations. In this procedure we restrict ourselves to the conditions \eqref{832mod} and \eqref{834mod}, since the remaining conditions \eqref{831mod} and \eqref{833mod} can be derived via the exact same techniques. We first derive the conditions \eqref{832mod} for the tangential component of $H$. At some arbitrary but fixed time $t \in [0,T]$ we pick an arbitrary point $x \in \partial S(t)$ and consider it the origin $x=(0,0,0)^T$ of a local coordinate system with axes $\theta$, $\zeta$ and $\eta$, where $\zeta$ is orthogonal to $\partial S(t)$ at $x$ while $\theta$ and $\eta$ are tangential to $\partial S(t)$ at $x$. We point out that in this construction the axis $\zeta$ is determined uniquely (except for its orientation). The axis $\eta$ may be chosen as an arbitrary axis intersecting $\zeta$ orthogonally in $x$ and subsequently the axis $\theta$ is obtained as the unique axis intersecting both $\zeta$ and $\eta$ orthogonally in $x$. Further, we introduce the outer unit normal vector $\text{n}_x=(0,1,0)^T$ on $\partial S(t)$ at $x$ and the unit vector $\text{n}_x'=(0,0,1)^T$, normal to the $\theta \zeta$-plane. Due to the smoothness assumptions \eqref{mathassumption1} on $\partial S(t)$ there exists some small $\Delta s > 0$ and a twice continuously differentiable function
\begin{align}
\phi: (-\Delta s,\Delta s)\rightarrow \mathbb{R},\quad \quad \phi (0)= \phi'(0)=0, \nonumber
\end{align}
such that the intersection of a small open neighborhood of $x=(0,0,0)^T$ in $\partial S(t)$ with the $\theta \zeta$-plane can be expressed as the set
\begin{align}
\left\lbrace \gamma (\theta):= \left(
\begin{array}{c}
\theta\\
\phi(\theta)\\
0\\
\end{array}
\right):\ \theta \in (-\Delta s, \Delta s) \right\rbrace. \label{curve}
\end{align}
Without loss of generality we assume points below the curve $\gamma$ defined by \eqref{curve} (i.e.\ points $(\theta, \zeta, 0)^T \in (-\Delta s, \Delta s) \times \mathbb{R} \times \{0\}$ with $\zeta < \phi(\theta)$) to lie inside of $S(t)$, while the points above $\gamma$ (i.e.\ points  $(\theta, \zeta, 0)^T \in (-\Delta s, \Delta s) \times \mathbb{R} \times \{0\}$ with $\zeta \geq \phi(\theta)$) are assumed to lie outside of $S(t)$. Shifting the curve $\gamma$ upwards as well as downwards along the axis $\zeta$ over a small distance $\Delta l > 0$, respectively, we introduce a curved rectangle $\Delta F$, enclosed by the four edges
\begin{align}
S_1 &= \left\lbrace \gamma _1(\zeta):= (\Delta s, \phi(\Delta s) + \zeta, 0)^T:\ \zeta \in (-\Delta l, \Delta l) \right\rbrace, \label{p1} \\
S_2 &= \left\lbrace \gamma_2(\theta):= (-\theta, \phi(-\theta) + \Delta l, 0)^T:\ \theta \in (-\Delta s,\Delta s) \right\rbrace, \label{p2} \\
S_3 &= \left\lbrace \gamma _3(\zeta):= (-\Delta s, \phi(-\Delta s) - \zeta, 0)^T:\ \zeta \in (-\Delta l,\Delta l) \right\rbrace, \label{p3} \\
S_4 &= \left\lbrace \gamma_4(\theta):= (\theta, \phi(\theta) - \Delta l, 0)^T:\ \theta \in (-\Delta s,\Delta s) \right\rbrace, \label{p4}
\end{align}
in the $\theta \zeta$-plane, cf.\ Figure \ref{curvedrectangle}. We multiply Ampère's law in the fluid domain (see \eqref{293}) and in the solid domain (see \eqref{1004}) by the vector $\text{n}_x'$ - which is normal to $\Delta F$ - and integrate the result over $\Delta F$. This yields the identity

\begin{align}
&\int_{S_1} H \cdot ds + \int_{S_2} H \cdot ds + \int_{S_3} H \cdot ds + \int_{S_4} H \cdot ds \nonumber \\
=& \int_{\partial \Delta F} H \cdot ds =\int_{\Delta F} (\operatorname{curl} H) \cdot \text{n}_x'\ dA = \int_{\Delta F} \left( \partial _t D + j + J \right) \cdot \text{n}_x'\ dA, \label{maxwellfaradayint}
\end{align}
where, for the sake of readability, we neglect the argument $t$ in the notation of the involved functions. Using the parametrizations \eqref{p1}--\eqref{p4} we can rearrange this equation by rewriting the integrals on the left-hand side. Using further the boundedness assumptions \eqref{mathassumption3} on $H$, $D$, $J$ as well as the $C^2$-regularity of $\phi$, we then deduce the inequality
\begin{align}
\left| \tau_x \cdot H\left(\begin{pmatrix} 0\\ \Delta l \\0\end{pmatrix} \right) - \tau_x \cdot H\left(\begin{pmatrix}0\\ - \Delta l\\ 0 \end{pmatrix}\right) - \frac{1}{2\Delta s} K(\Delta F) \right| \leq c(\Delta l + \Delta s), \label{wefujwe09uuj09}
\end{align}
where $\tau_x := \text{n}_x' \times \text{n}_x = (-1,0,0)^T$, $c>0$ is a constant independent of $\Delta l$ and $\Delta s$ and
\begin{align}
K(\Delta F) := \int_{\Delta F} j \cdot \text{n}_x'\ dA \label{Kdef}
\end{align}
denotes the current in the curved rectangle $\Delta F$. For the details of the derivation of this estimate we refer to Section \ref{etangential} in the appendix. We let first $\Delta l$ and subsequently $\Delta s$ tend to zero. Then $\frac{1}{2\Delta s}K(\Delta F)$ converges to the surface current density (i.e.\ the current per unit length)
\begin{align}
k = k_x := \lim_{\Delta s \rightarrow 0} \lim_{\Delta l \rightarrow 0} \frac{1}{2\Delta s} K(\Delta F) \label{surfacecurrentdensity}
\end{align}
in the origin $x=(0,0,0)^T$ of the local coordinate system and we infer that
\begin{align}
\left( H^f \left( \begin{pmatrix} 0\\ 0 \\0\end{pmatrix}\right) - H^s \left( \begin{pmatrix} 0\\ 0 \\0\end{pmatrix} \right) \right) \cdot \left( \text{n}_x' \times \text{n}_x \right) = \left( H^f \left( \begin{pmatrix} 0\\ 0 \\0\end{pmatrix}\right) - H^s \left( \begin{pmatrix} 0\\ 0 \\0\end{pmatrix} \right) \right) \cdot \tau_x = k_x. \nonumber
\end{align}
Due to the arbitrary choice of the axis $\eta$ and hence the vector $\text{n}_x'$ as well as the arbitrary choice of the origin $x = (0,0,0)^T \in \partial S(t)$ of the local coordinate system we may drop the index $x$ and infer the interface condition
\begin{equation}
\left( H^f - H^S \right) \times \text{n} = k \quad \text{on } \partial S(t). \label{835}
\end{equation}
Here, we assume that $k=0$ on $\partial S(t)$, i.e.\ the right-hand side of the identity \eqref{835} is zero and consequently we infer the second equation for $H$ in \eqref{832mod}. This is in accordance with the literature, according to which the surface current density $k$ vanishes on the surfaces of most materials. More specifically, for example in \cite[Section 9.4.2]{griffiths}, it is stated that the equality \eqref{835} with zero right-hand side holds true on the interfaces between Ohmic conductors, i.e.\ conductors (with finite conductivity $\sigma < \infty$) in which Ohm's law holds true. Indeed, provided that Ohm's law is also satisfied on the interface itself, the identity $k=0$ follows directly from the definition of $k$ in \eqref{surfacecurrentdensity} and the boundedness assumptions \eqref{mathassumption3} on $E$, $u$ and $B$. However, we point out that the applicability of Ohm's law on the interface poses an additional assumption, the validity of which does not seem to be generally accepted. 
It remains an open problem to mathematically justify $k=0$ in the case that Ohm's law is not satisfied at the interface. 

In general, $k$ does not necessarily need to vanish. From the mathematical point of view, $k$ may take values different from zero if $j$ becomes infinite on the considered surface, which can be expressed mathematically via the use of a Dirac delta distribution. Also according to the physical literature there exist materials, such as superconductors, on the surfaces of which $k$ takes non-zero values, cf.\ \cite[Chapter 11]{greinere}. In the condition for $H$ on $\partial \Omega$ in \eqref{832mod}, which is derived by the same arguments as the identity \eqref{835}, we thus refrain from the restrictive assumption $k=0$. Indeed, while on $\partial S(t)$ we make this assumption in order for the interface condition to match the one used in the model \ref{2919}--\eqref{2908} for the analytical work in the article \cite{compressiblepaper}, on $\partial \Omega$ we regard it more appropriate to allow $k$ to take non-zero values.

\begin{figure}
\centering
\includegraphics[scale=0.3]{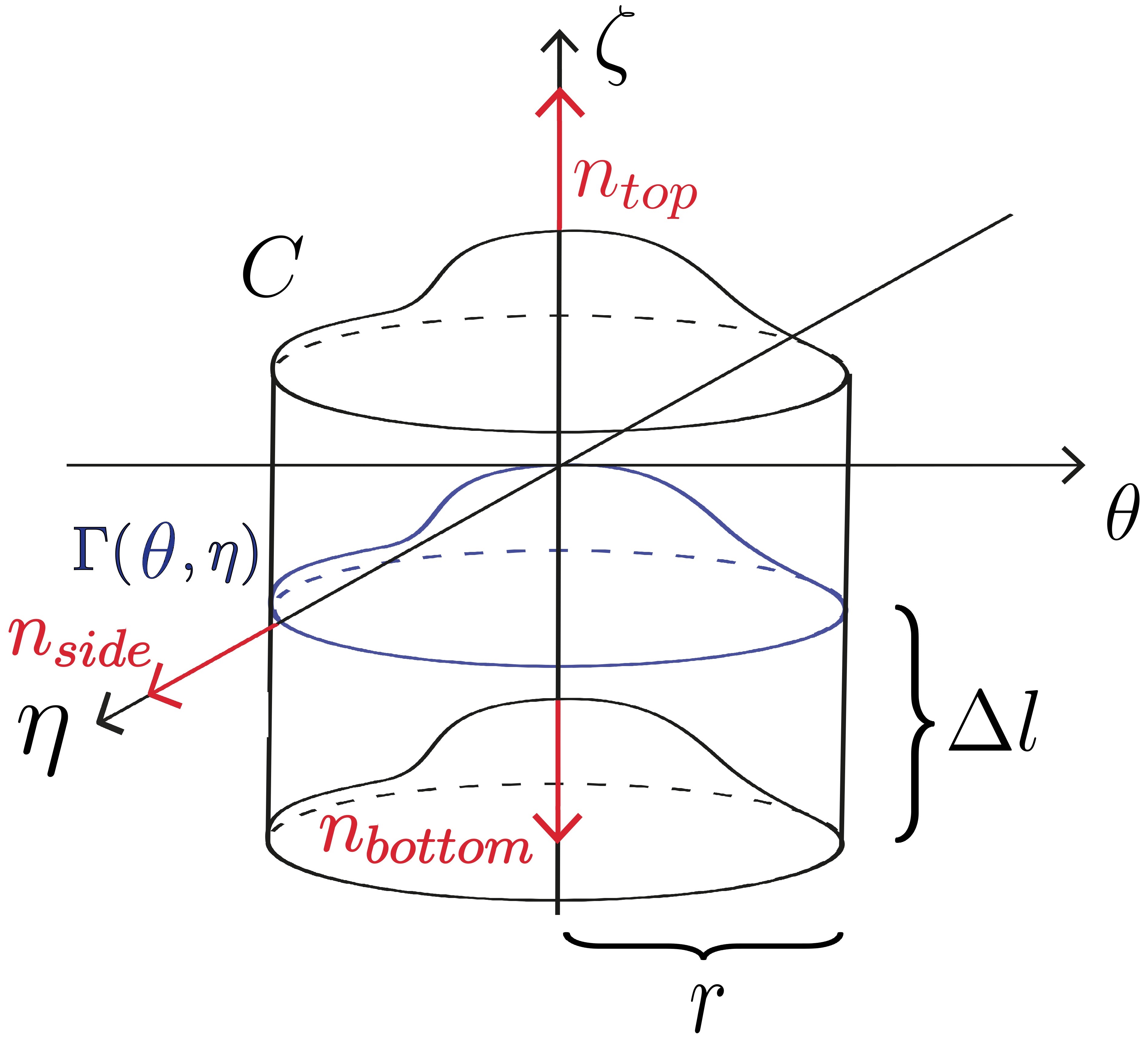} 
\hspace*{0.3cm}
\caption{The cylinder $C$ with curved bases.} \label{curvedcylinder}
\end{figure}

Next we deduce the conditions \eqref{834mod} for the normal component of $D$. Again we choose an arbitrary point $x \in \partial S(t)$ as the origin $x=(0,0,0)^T$ of a local coordinate system with axes $\theta$, $\zeta$ and $\eta$, where $\zeta$ is orthogonal to $\partial S(t)$ at $x$ while $\theta$ and $\eta$ are tangential to $\partial S(t)$ at $x$. We denote by $B_r(0)$ the $2$-dimensional open ball with radius $r>0$ centered at $x=(0,0,0)^T$ in the $\theta \eta$-plane of this local coordinate system. Due to the smoothness assumptions \eqref{mathassumption1} on $S(t)$ we can choose $r$ sufficiently small and find some twice continuously differentiable function
\begin{align}
\Phi :B_r(0) \rightarrow \mathbb{R},\quad \quad \Phi(0,0) = \partial_i \Phi (0,0)=0 \quad \text{for } i =1,2, \nonumber
\end{align}
where $\partial_i \Phi$ denotes the derivative of $\Phi$ with respect to the $i$-th variable, such that a small open neighborhood of $x$ in $\partial S(t)$ can be written as the set
\begin{align}
\left\lbrace \Gamma (\theta, \eta) := \begin{pmatrix} \theta \\ \Phi (\theta, \eta) \\ \eta \end{pmatrix}:\ (\theta , \eta) \in B_r(0) \right\rbrace. \label{surface}
\end{align}
Without loss of generality we assume points below the surface $\Gamma$ defined by \eqref{surface} (i.e.\ points $(\theta, \zeta, \eta)^T$ with $(\theta, \eta) \in B_r(0)$ and $\zeta < \Phi(\theta, \eta)$) to lie inside of $S(t)$ while the points above $\Gamma$ (i.e.\ points $(\theta, \zeta, \eta)^T$ with $(\theta, \eta) \in B_r(0)$ and $\zeta > \Phi(\theta, \eta)$) are assumed to lie outside of $S(t)$. By shifting the surface $\Gamma$ upwards and downwards along the axis $\zeta$ over a small distance $\Delta l > 0$ we define a cylinder $C$ with the curved bases
\begin{align}
\Delta _{\text{top}} &:= \left\lbrace \Gamma_{\text{top}}(s,\alpha) := \begin{pmatrix} s \cos(\alpha) \\ \Phi (s \cos (\alpha), s \sin (\alpha)) + \Delta l \\ s \sin(\alpha) \end{pmatrix}:\ 0 \leq s \leq r,\ 0 \leq \alpha \leq 2 \pi \right\rbrace, \label{f43j0j4f04f} \\
\Delta _{\text{bottom}} &:= \left\lbrace \Gamma_{\text{bottom}}(s,\alpha) := \begin{pmatrix} s \cos(-\alpha)\\ \Phi (s \cos (-\alpha), s \sin (-\alpha)) - \Delta l \\ s \sin(-\alpha)  \end{pmatrix}:\ 0 \leq s \leq r,\ 0 \leq \alpha \leq 2 \pi \right\rbrace \label{d3f3j0j4fjj}
\end{align}
and the lateral area
\begin{align}
\Delta _{\text{side}} &:= \left\lbrace \Gamma_{\text{side}}(\alpha,h) := \begin{pmatrix} r\cos(\alpha) \\ \Phi (r \cos (\alpha), r \sin (\alpha)) + h \\ r\sin(\alpha) \end{pmatrix}:\ 0 \leq \alpha < 2\pi,\ -\Delta l \leq h \leq \Delta l \right\rbrace, \label{dr34f304f0i3}
\end{align}
cf.\ Figure \ref{curvedcylinder}. Moreover, we denote by $\text{n}_C$ the outer unit normal vector on $\partial C$, which we split into the outer unit normal vector $\text{n}_{\text{top}}$ on $\Delta_{\text{top}}$, the outer unit normal vector $\text{n}_{\text{bottom}}$ on $\Delta_{\text{bottom}}$ and the outer unit normal vector $\text{n}_{\text{side}}$ on $\Delta_{\text{side}}$.

We integrate Gauss's law in the fluid domain (cf.\ \eqref{290}) and in the solid domain (cf.\ \eqref{1001}) over the cylinder $C$. This leads to the identity
\begin{equation}
\int_{\Delta _\text{top}} D \cdot \text{n}_{\text{top}}\ dA + \int_{\Delta _\text{bottom}} D \cdot \text{n}_\text{bottom}\ dA + \int_{\Delta _\text{side}} D \cdot \text{n}_\text{side}\ dA = \int_{\partial C} D \cdot \text{n}_C\ dA = \int_{C} \text{div}D\ dx = \int_C \rho_c\ dx. \label{rfrwweügiii}
\end{equation}
This equation can be rearranged by rewriting the integrals on the left-hand side under usage of the parametrizations \eqref{f43j0j4f04f}--\eqref{dr34f304f0i3}. Then, under exploitation of the assumed boundedness of $B$ (see \eqref{mathassumption3}) as well as the $C^2$-regularity of $\Phi$, we deduce the estimate
\begin{align}
\left| \text{n}_x \cdot \left[ D\left( \begin{pmatrix} 0 \\ -\Delta l \\ 0 \end{pmatrix}\right) - D\left( \begin{pmatrix} 0 \\ \Delta l \\ 0 \end{pmatrix}\right) \right] - \frac{1}{\pi r^2} W(C) \right| \leq c (r + \Delta l) \label{boundaryestimate}
\end{align}
for the outer unit normal vector $\text{n}_x = (0,1,0)^T$ on $\partial S(t)$ in the origin $x=(0,0,0)^T$ of the local coordinate system, a constant $c>0$ independent of $r$ and $\Delta l$ and the electric charge
\begin{align}
W(\Delta F) := \int_C \rho_c \ dX \label{Wdef}
\end{align}

of the cylinder $C$. More details on the derivation of this estimate are given in Section \ref{bnormal} in the appendix. We let $r$ and $\Delta l$ tend to zero in the inequality \eqref{boundaryestimate}. This limit passage turns the quantity $\frac{1}{\pi r^2}W(C)$ into the surface charge density
\begin{align}
\omega = \omega_x := \lim_{r \rightarrow 0} \lim_{\Delta l \rightarrow 0} \frac{1}{\pi r^2} W(C) \nonumber
\end{align}

in the origin $x=(0,0,0)^T$ of the local coordinate system. Since $x \in \partial S(t)$ was chosen arbitrarily we thus infer the second condition for $D$ in \eqref{834mod}. The first condition follows by the same arguments.

\section{Magnetohydrodynamic approximation via nondimensionalization} \label{non-dimensionalization}

In this section we carry out a nondimensionalization of various equations from the system introduced in Section \ref{originalmodel} in order to single out several negligibly small terms. We will then neglect these terms from the system in order to obtain a simplified model. In particular, in the fluid domain this simplification constitutes the classical magnetohydrodynamic approximation, an alternative derivation of which can be found in \cite{jiliang,jiliang2}. For any physical quantity $a$ we denote by $\overline{a}> 0$ an associated characteristic scale. Introducing the dimensionless variables
\begin{align}
t' &:= \frac{t}{\overline{t}},\quad \quad T' := \frac{T}{\overline{t}},\quad \quad x' := \frac{x}{\overline{x}} \label{fewofokpwejgkopew}
\end{align}
we may then define nondimensionalized versions of our several domains,
\begin{equation}
\Omega' := \frac{1}{\overline{x}}\Omega,\quad \quad S'(t') := \frac{1}{\overline{x}} S(t'\overline{t}) = \frac{1}{\overline{x}} S(t),\quad \quad F'(t') := \Omega' \setminus \overline{S'(t')} = \frac{1}{\overline{x}}F(t) \nonumber
\end{equation}
and
\begin{equation}
Q':=(0,T')\times \Omega',\quad \quad {Q^s}' := \left\lbrace (t',x') \in Q' :\ x' \in S'(t') \right\rbrace,\quad \quad {Q^f}' := \left\lbrace (t',x') \in Q' :\ x' \in F'(t') \right\rbrace. \nonumber
\end{equation}
We remark that
\begin{equation}
(t,x) \in Q \Leftrightarrow (t',x') \in Q',\quad \quad (t,x) \in Q^S \Leftrightarrow (t',x') \in {Q^S}',\quad \quad (t,x) \in Q^f \Leftrightarrow (t',x') \in {Q^f}'. \nonumber
\end{equation}
Moreover, for any $(t',x') \in Q'$ we introduce the dimensionless mechanical quantities
\begin{align}
\rho'&(t',x') := \frac{\rho (t,x)}{\overline{\rho}},\ \ \ u'(t',x') := \frac{u(t,x)}{\overline{u}},\ \ \ p'(t',x') := \frac{p(t,x)}{\overline{p}},\ \ \ g'(t',x') := \frac{g(t,x)}{\overline{g}}, \label{uu}
\end{align}
and for all $(t',x') \in (0,T')\times \mathbb{R}^3$ we introduce the dimensionless electromagnetic quantities
\begin{align}
D'(t'&,x') := \frac{D(t,x)}{\overline{D}},\ \ \ E'(t',x') := \frac{E(t,x)}{\overline{E}},\ \ \ B'(t',x') := \frac{B(t,x)}{\overline{B}},\ \ \ H'(t',x') := \frac{H(t,x)}{\overline{H}},\label{vv} \\
&\quad \quad \quad \ \ \rho _c'(t',x') := \frac{\rho _c(t,x)}{\overline{\rho}_c},\ \ \ j'(t',x') := \frac{j(t,x)}{\overline{j}},\ \ \ J'(t',x') := \frac{J(t,x)}{\overline{j}}. \label{dwe0fiqkwf0i}
\end{align}

\subsection{The Maxwell system} \label{mhdapprox}

In the classical magnetohydrodynamic approximation, the Maxwell system in the fluid domain is simplified by dropping the quantity $\partial_t D$ in Ampère's law. The purpose of this section is to justify this simplification. Furthermore, proceeding similarly in the solid region, we carry out the same reduction in Ampère's law in the solid domain. We start by nondimensionalizing the Maxwell-Faraday equation \eqref{292} in the fluid and the solid domain. Expressing the magnetic induction $B$ and the electric field $E$ through the dimensionless variables introduced in \eqref{fewofokpwejgkopew} and \eqref{vv} and making use of the chain rule, the equation \eqref{292} becomes
\begin{equation}
\frac{\overline{B}}{\overline{t}} \partial _{t'} B'(t',x') + \frac{\overline{E}}{\overline{x}} \nabla _{x'} \times E'(t',x') = 0\quad \quad \text{in } {Q^f}'\ \text{and}\ {Q^s}'. \nonumber
\end{equation}
In this relation we assume the two quantities $\partial _{t'} B'$ and $\nabla _{x'} \times E'$ to be equally significant, meaning that the coefficients in front of these terms have to coincide,
\begin{equation}
\frac{\overline{B}}{\overline{t}} = \frac{\overline{E}}{\overline{x}}. \label{an}
\end{equation}
This relation, together with the linear relations \eqref{304} and the decomposition \eqref{muepsilon} of $\mu$ and $\epsilon$, allows us to express Ampère's law \eqref{293} in the fluid domain and \eqref{1004} in the solid domain as
\begin{align}
    \nabla_{x'} \times B'\left(t',x' \right) &= \left\{
                \begin{array}{ll}
                  \frac{\mu_0\mu_f^r \overline{x} \epsilon_0 \epsilon_f^r \overline{E}}{\overline{t}\ \overline{B}} \partial_{t'} E' \left(t',x' \right) + \frac{\mu^f \overline{x}\ \overline{j}}{\overline{B}}j'\left(t',x' \right) + \frac{\mu^f \overline{x}\ \overline{j}}{\overline{B}}J'\left(t',x' \right) \ \ \ &\text{in } Q^f,\\
                  \frac{\mu_0 \overline{x} \epsilon_0 \overline{E}}{\overline{t}\ \overline{B}} \partial_{t'} E' \left(t',x' \right) &\text{in } Q^s.
                \end{array}
              \right. \label{nondimamp}
\end{align}
In the fluid domain, we assume the quantities $j'$ and $\nabla_{x'} \times B'$ to be equally significant, meaning that
\begin{equation}
\overline{B} = \mu^f \overline{x} \overline{j}, \label{ar}
\end{equation}
Denoting by $c>0$ the speed of light, which is known to satisfy
\begin{align}
c=(\mu_0 \epsilon_0)^{-\frac{1}{2}}, \label{lightspeed}
\end{align}
we further assume that
\begin{equation}
\overline{u} = \frac{\overline{x}}{\overline{t}},\quad \quad \overline{u} \ll c,\quad \quad \mu_r^f \epsilon_r^f \approx 1. \label{ao}
\end{equation}
We point out that the latter of the assumptions in \eqref{ao} is reasonable for isotropic liquid electrical conductors (i.e. for example electrolytes, molten salts and liquid metals), cf. \cite[Section II.1]{moreau}. Combining the relation \eqref{an} with the first assumption in \eqref{ao}, we infer that
\begin{equation}
\frac{\overline{E}}{\overline{B}} = \overline{u}. \label{419}
\end{equation}
This, together with the relations \eqref{ar}--\eqref{ao}, allows us to simplify the nondimensionalized version \eqref{nondimamp} of Ampère's law to
\begin{align}
    \nabla_{x'} \times B'\left(t',x' \right) &= \left\{
                \begin{array}{ll}
                  \frac{\overline{u}^2}{c^2} \partial_{t'} E' \left(t',x' \right) + j'\left(t',x' \right) + J'\left(t',x' \right) \ \ \ &\text{in } Q^f,\\
                  \frac{\overline{u}^2}{c^2} \partial_{t'} E' \left(t',x' \right) &\text{in } Q^s.
                \end{array}
              \right. \label{redamp}
\end{align}
Here, the term $\frac{\overline{u}^2}{c^2}\partial_{t'}E'$ is negligibly small according to the second assumption in \eqref{ao}. Formally, we may hence neglect it, which reduces the identity \eqref{redamp} to
\begin{align}
    \nabla_{x'} \times B'\left(t',x' \right) &= \left\{
                \begin{array}{ll}
                  j'\left(t',x' \right) + J'\left(t',x' \right) \ \ \ &\text{in } Q^f,\\
                  0 &\text{in } Q^s.
                \end{array}
              \right. \label{420}
\end{align}
The desired reduced version of Ampère's law is then obtained by converting this identity back into a dimensional form,
\begin{align}
    \nabla \times H\left(t,x \right) &= \left\{
                \begin{array}{ll}
                  j\left(t,x \right) + J\left(t,x \right) \ \ \ &\text{in } Q^f,\\
                  0 &\text{in } Q^s.
                \end{array}
              \right. \label{421}
\end{align}

\subsection{The Navier-Stokes system}

In the Navier-Stokes system we nondimensionalize the momentum equation \eqref{1.2} with the aim of simplifying the Lorentz force $\rho _c E + (j+J) \times B$ to the expression $\frac{1}{\mu} \operatorname{curl}B \times B$ as it is common in the magnetohydrodynamic approximation. A straight forward calculation under exploitation of the relations \eqref{fewofokpwejgkopew}, \eqref{uu}, \eqref{vv} and \eqref{dwe0fiqkwf0i} allows us to express the momentum equation \eqref{1.2} in the form
\begin{align}
&\frac{\overline{\rho}\ \overline{u}}{\overline{t}} \partial _{t'} \left(\rho '(t',x') u'(t',x')\right) + \frac{\overline{\rho}\ \overline{u}^2}{\overline{x}} \nabla_{x'} \cdot \left( \rho '(t',x') u'(t',x') \otimes u'(t',x')\right) + \frac{\overline{p}}{\overline{x}}\nabla_{x'}p'(t',x') \nonumber \\
=& \frac{2\nu \overline{u}} {\overline{x}^2}\nabla _{x'} \cdot \mathbb{D}_{x'}\left( u'(t',x') \right) + \frac{\lambda \overline{u}}{\overline{x}^2} \nabla_{x'} \cdot \left( \operatorname{id} \left( \nabla_{x'} \cdot u'\left(t',x' \right)\right) \right) + \overline{\rho}\ \overline{g} \rho '(t',x') g(t,x) + \overline{\rho}_c \overline{E}\rho_c'(t',x')E'(t',x') \nonumber \\
&+ \overline{j}\ \overline{B}\left(j'(t',x') + J'(t',x') \right) \times B'(t',x')\quad \quad \text{in } Q^{f'}. \label{5181}
\end{align}
We proceed by taking the divergence of Ohm's law \eqref{306} in the fluid domain, which, under exploitation of Gauss's law \eqref{290} and the linear relations \eqref{304}, yields the identity
\begin{align}
\frac{1}{\sigma}\nabla \cdot j(t,x) - \nabla \cdot \left(u(t,x)\times B(t,x) \right) = \frac{1}{\epsilon^f} \nabla \cdot D(t,x) = \frac{\rho_c(t,x)}{\epsilon^f} \quad \quad \text{in } Q^f. \nonumber
\end{align}
Nondimensionalizing this relation we obtain
\begin{align}
\frac{\overline{j}}{\sigma \overline{x}} \nabla_{x'} \cdot j'(t',x') - \frac{\overline{u}\overline{B}}{\overline{x}} \nabla_{x'} \cdot \left(u'(t',x') \times B'(t',x') \right) = \frac{\overline{\rho}_c }{\epsilon^f}\rho_c'(t',x') \quad \quad \text{in } {Q^f}'.\nonumber
\end{align}
Here the first term on the left-hand side is equal to zero since $\nabla_{x'}\cdot j' = 0$ due to Ampère's law \eqref{420} and the fact that (by the assumptions on $J$ in \eqref{assumptionsJ}) $J'$ is divergence-free. We infer that
\begin{equation}
\frac{\overline{\rho}_c \overline{x}}{\epsilon^f \overline{u}  \overline{B}} \rho '_c(t',x') = - \nabla _{x'} \cdot \left( u'(t',x') \times B'(t',x') \right)\quad \text{in } Q^{f'}. \nonumber
\end{equation}
In this identity we assume the quantities $\rho '_c$ and $-\nabla _{x'} \cdot \left( u' \times B' \right)$ to be equally significant, meaning that
\begin{equation}
\overline{\rho}_c = \frac{\epsilon^f \overline{u}\overline{B}}{\overline{x}}. \label{as}
\end{equation}
This yields the equation
\begin{align}
\overline{\rho}_c \overline{E} = \frac{\epsilon^f \overline{u}\overline{B}}{\overline{x}} \overline{E} = \frac{\epsilon^f \overline{B}^2\overline{u}^2}{\overline{x}} = \frac{\epsilon _0 \epsilon _r^f \mu _0 \mu _r^f \overline{x} \overline{j}\ \overline{B}\overline{u}^2}{\overline{x}} \approx \frac{\overline{u}^2}{c^2} \overline{j}\ \overline{B}, \nonumber
\end{align}
where we used the relation \eqref{419} for the second identity, the relations \eqref{muepsilon} and \eqref{ar} for the third identity and the relation \eqref{lightspeed} as well as the third assumption in \eqref{ao} for the last identity. For the (nondimensionalized) Lorentz force this means that
\begin{align}
&\overline{\rho}_c\ \overline{E} \rho _c'(t',x') E'(t',x') + \overline{j}\ \overline{B}\left(j'(t',x') + J'(t',x') \right) \times B'(t',x') \nonumber \\
\approx& \frac{\overline{u}^2}{c^2} \overline{j}\ \overline{B} \rho _c'(t',x') E'(t',x') + \overline{j}\ \overline{B}\left(j'(t',x') + J'(t',x') \right) \times B'(t',x')\quad \quad \text{in } Q^{f'}. \nonumber
\end{align}
Due to the second assumption in \eqref{ao} we see that the first term on the right-hand side of this relation is negligibly small compared to the second one. Hence, formally, we may neglect it and the nondimensionalized momentum equation \eqref{5181} reduces to
\begin{align}
&\frac{\overline{\rho}\ \overline{u}}{\overline{t}} \partial _{t'} \left(\rho '(t',x') u'(t',x')\right) + \frac{\overline{\rho}\ \overline{u}^2}{\overline{x}} \nabla_{x'} \cdot \left( \rho '(t',x') u'(t',x') \otimes u'(t',x')\right) + \frac{\overline{p}}{\overline{x}}\nabla_{x'}p'(t',x') \nonumber \\
=& \frac{2\nu \overline{u}} {\overline{x}^2}\nabla _{x'} \cdot \mathbb{D}_{x'}\left( u'(t',x') \right) + \frac{\lambda \overline{u}}{\overline{x}^2} \nabla_{x'} \cdot \left( \operatorname{id} \left( \nabla_{x'} \cdot u'\left(t',x' \right)\right) \right) + \overline{\rho}\ \overline{g} \rho '(t',x') g(t,x) \nonumber \\
&+ \overline{j}\ \overline{B}\left(j'(t',x') + J'(t',x') \right) \times B'(t',x')\quad \quad \text{in } Q^{f'}. \nonumber
\end{align}
Transforming this equation back into a dimensional form and exploiting Ampère's law \eqref{421} and the linear relation \eqref{304} to rewrite the remaining part of the Lorentz force, we obtain the desired simplified momentum equation
\begin{align}
&\partial _{t} \left( \rho (t,x) u(t,x)\right) + \nabla \cdot \left( \rho(t,x) u(t,x) \otimes u(t,x) \right) + \nabla p(t,x) \nonumber \\
=& 2\nu \nabla \cdot \mathbb{D} \left( u(t,x) \right) + \lambda \nabla \cdot \left( \operatorname{id} \left( \nabla \cdot u\left(t,x \right)\right) \right) + \rho(t,x) g(t,x) + \frac{1}{\mu} \left( \nabla \times B(t,x) \right) \times B(t,x)\quad \quad \text{in } Q^f. \label{1.3}
\end{align}

\section{Summary of the derived system} \label{finalsystem}

In the following we present a summary of the system derived in the previous sections. The mechanical part of this system coincides with the mechanical part \eqref{1.1}--\eqref{1.9}, \eqref{noslip} of the original system except for the momentum equation, which has been simplified in the course of the magnetohydrodynamic approximation, cf.\ \eqref{1.3}.
In the electromagnetic part of the derived system, the Maxwell-Faraday equation, Gauss's law and Gauss's law for magnetism in the fluid domain are adopted directly from the original model, cf.\ \eqref{292}--\eqref{291}. In the solid domain, these equations have been adjusted to the assumption of the solid being insulating in \eqref{1003}--\eqref{1002}. Ampère's law has been adjusted to the same assumption and has undergone, in both the fluid and the solid domain, the magnetohydrodynamic approximation; the resulting equation is stated in \eqref{421}. The Maxwell system in the exterior domain in our derived model consists, in accordance with the assumption of the exterior domain being a perfect conductor, only of Ampère's law \eqref{ampext} and Gauss's law \eqref{gaussext}, since the remaining equations become superfluous due to the trivial relations \eqref{vanishingfields}. Ohm's law in the derived system keeps its form \eqref{306} from the original model, however it can be omitted in the exterior domain because of the trivial relations \eqref{vanishingfields}. The constitutive relations \eqref{300} from the original model reduce according to the trivial relations \eqref{vanishingfields} in the exterior domain while they take the form \eqref{304} in the fluid and the solid domain. Finally, our derived model also includes the interface conditions \eqref{831mod}--\eqref{834mod} for the electromagnetic fields.

Before we present the derived system in its complete form we recall the conditions which we obtained through the scaling assumptions we used for the magnetohydrodynamic approximation in Section \ref{non-dimensionalization}. These conditions consist of the relations
\begin{align}
\frac{\overline{B}}{\overline{t}} = \frac{\overline{E}}{\overline{x}}, \quad \quad \overline{B} = \mu^f \overline{x} \overline{j}, \quad \quad \overline{u} = \frac{\overline{x}}{\overline{t}}, \quad \quad \overline{u} \ll c, \quad \quad \mu_r^f \epsilon_r^f \approx 1, \quad \quad \overline{\rho}_c = \frac{\epsilon^f \overline{B}\overline{u}}{\overline{x}} \nonumber
\end{align}
for the characteristic scales $\overline{a}>0$ of the physical quantities $a$ in our model, cf.\ \eqref{an}, \eqref{ar}, \eqref{ao} and \eqref{as}.

The full system we derived under these conditions for the modeling of the motion of several insulating rigid bodies through an electrically conducting diamagnetic dielectric viscous non-homogeneous and compressible fluid surrounded by a perfect conductor is composed of the Maxwell equations
\begin{align}
\text{curl} H &= \left\{
                \begin{array}{ll}
                  j + J \ \ \ &\text{in } Q^f,\\
                  0 &\text{in } Q^s, \\
                  \partial_t D &\text{in } Q^{\operatorname{ext}},
                \end{array}
              \right. \label{w3fejwejojjo} \\
\partial_t B + \operatorname{curl} E &= 0 \quad \quad \quad \quad \quad \text{in } Q^f \text{ and } Q^s, \label{ewofjweojf} \\
\operatorname{div} D &= \left\{
                \begin{array}{ll}
                  \rho_c \quad \quad \ \ &\text{in } Q^f \text{ and } Q^{\operatorname{ext}},\\
                  0 &\text{in } Q^s,
                \end{array}
              \right. \label{ewoejfsejf} \\
\operatorname{div} B &= 0 \quad \quad \quad \quad \quad \text{in } Q^f \text{ and } Q^s, \label{jweo0jv0wsjd0}
\end{align}
the compressible Navier-Stokes equations
\begin{align}
\partial _t \rho + \operatorname{div} (\rho u) =& 0 \quad \quad \quad \quad \quad \quad \quad \quad \quad \quad \quad \quad \ \text{in } Q^f, \label{526mod} \\
\partial _{t} (\rho u) + \text{div} (\rho u \otimes u) + \nabla p =& \text{div} \mathbb{T} + \rho g + \frac{1}{\mu} \text{curl}B\times B\quad \quad \text{in } Q^f, \label{527mod}
\end{align}
the balances of linear and angular momentum
\begin{align}
m^i \frac{d}{dt} V^i(t) =& \frac{d}{dt} \int_{S^i(t)} \rho u\ dx = \int_{\partial S^i(t)} [\mathbb{T} - p\text{ id}] \text{n}\ dA + \int _{S^i(t)} \rho g\ dx, \quad t \in [0,T], \label{528mod} \\
\frac{d}{dt}\left( \mathbb{J}^i(t)w^i(t) \right) =& \frac{d}{dt} \int _{S^i(t)} \rho \left(x - X^i\right) \times u\ dx \nonumber \\
=& \int_{\partial S^i(t)} (x - X^i) \times [\mathbb{T} - p\ \text{id} ] \text{n}\ dA + \int _{S^i(t)} \rho \left(x - X^i\right) \times g\ dx, \quad 
t \in [0,T], \label{529mod}
\end{align}
for $i=1,...,N$, in combination with the trivial relations
\begin{align}
B = E = j = 0\quad \quad \text{in } Q^{\text{ext}}, \label{exteriorfinal}
\end{align}
the relations
\begin{align}
j =& \sigma (E + u \times B) \quad \text{in } Q^f \text{ and } Q^s,  \quad \quad \sigma = \left\{
                \begin{array}{ll}
                  \sigma ^f > 0 \ \ \ &\text{in } Q^f,\\
                  \sigma ^s = 0 &\text{in } Q^s,
                \end{array}
              \right. \label{524} \\
B =& \mu H \quad \quad \text{in } Q^f\ \text{and } Q^s,\quad \quad \quad \quad \ D= \epsilon E \quad \quad \text{in } Q^f\ \text{and } Q^s,\label{523} \\
H =& -M \quad \ \text{in } Q^{\text{ext}},\quad \quad \quad \quad \quad \quad \quad D = P \quad \quad \ \ \text{in } Q^{\text{ext}}
\end{align}
and the boundary and interface conditions
\begin{align}
E^f(t) \times \text{n} &= 0 \quad \quad \ \text{on } \partial \Omega,\quad \quad &&\left(E^f(t) - E^s(t)\right) \times \text{n} = 0 \quad \quad \ \ \text{on } \partial S(t), \label{bic1} \\
\left(H^{\text{ext}}(t) - H^f(t)\right) \times \text{n} &= k(t) \quad \text{on } \partial \Omega,\quad \quad &&\left(H^f(t) - H^s(t)\right) \times \text{n} = 0 \quad \quad \ \text{on } \partial S(t), \label{bic2} \\
B^f(t) \cdot \text{n} &= 0\quad \quad \ \text{on } \partial \Omega,\quad \quad &&\ \ \left(B^f(t) - B^s(t)\right) \cdot \text{n} = 0 \quad \ \quad \text{on } \partial S(t), \label{bic3} \\
\left(D^{\text{ext}}(t) - D^f(t)\right) \cdot \text{n} &= \omega(t) \quad \text{on } \partial \Omega,\quad \quad &&\ \ \left(D^f(t) - D^s(t)\right) \cdot \text{n} = \omega(t) \quad \text{on } \partial S(t), \label{bic4} \\
u^f(t) &= 0\quad \ \quad \text{on } \partial \Omega,\quad \quad && \quad \quad \ \quad \ \ u^f(t) - u^s(t) = 0 \quad \ \quad \text{on } \partial S(t). \label{bic5}
\end{align}
In this system, the external force $J: (0,T) \times \mathbb{R}^3 \rightarrow \mathbb{R}^3$ in Ampère's law \eqref{w3fejwejojjo} in the fluid domain is assumed to satisfy
\begin{align}
\nabla \cdot J = 0 \quad \quad \text{in } Q^f,\quad \quad \quad \quad J = 0 \quad \quad \text{in } Q^s \text{ and } Q^{\text{ext}}. \label{4023}
\end{align}
The magnetic permeability $\mu$ and the dielectric permittivity $\epsilon$ in the relations \eqref{523} are given by
\begin{align}
 \quad \mu := \left\{
                \begin{array}{ll}
                  \mu ^f := \mu_0 \mu_r^f > 0 \ \ \ &\text{in } Q^f,\\
                  \mu ^s := \mu_0 > 0 &\text{in } Q^s,
                \end{array}
              \right. \quad \epsilon = \left\{
                \begin{array}{ll}
                  \epsilon ^f := \epsilon_0 \epsilon_r^f > 0 \ \ \ &\text{in } Q^f,\\
                  \epsilon ^s := \epsilon_0 > 0 &\text{in } Q^s,
                \end{array}
              \right. \nonumber
\end{align}
wherein $\mu_r^f$ and $\epsilon_r^f$ denote the relative permeability and the relative permittivity of the fluid. Finally, in the interface conditions \eqref{bic2}, \eqref{bic4} the quantities $k$ and $\omega$ denote the surface current density and the surface charge density, respectively, on $\partial \Omega$ and $\partial S(t)$.

The derived system \eqref{w3fejwejojjo}--\eqref{bic5} finds itself in an intermediate state between the general system \eqref{293}--\eqref{noslip} and the system \eqref{2919}--\eqref{2908} from the mathematical analysis, the derivation of which is the ultimate goal of the present article. The system \eqref{293}--\eqref{noslip} models the interaction between an electrically conducting fluid and an insulating rigid body in full generality. The equations \eqref{2919}--\eqref{2908}, as a reduced version of this system, are less general, however, they bear the advantage of admitting weak solutions as the authors were able to show in \cite{compressiblepaper}. The disadvantage of the latter system lies in the fact that some of its modifications in comparison to the system \eqref{293}--\eqref{noslip} are made for purely mathematical reasons. This, in turn, shows why the system \eqref{w3fejwejojjo}--\eqref{bic5} is interesting: It constitutes an intermediate result in the derivation of the system \eqref{2919}--\eqref{2908} from the system \eqref{293}--\eqref{noslip} containing only those modifications for which we have been able to provide - under certain assumptions - physical arguments. However, we point out that existence of weak solutions to the system \eqref{w3fejwejojjo}--\eqref{bic5} remains a challenging problem due to the jump of the magnetic permeability $\mu$ across the interface between the fluid and the solid in this system, which would require a new variational formulation and, probably, a new methodology.

We end this section by discussing the final adjustments which need to be made in order to turn the system \eqref{w3fejwejojjo}--\eqref{bic5} into the system \eqref{2919}--\eqref{2908}. In the latter system, as it is common practice in many mathematical works in magnetohydrodynamics (cf.\ for example \cite{blancducomet}), the Maxwell equations are only considered inside of the domain $\Omega$. This is due to the additional assumption of the linear relations \eqref{523} holding true also in the exterior domain,
\begin{align}
B = \mu H \quad \quad \text{in } Q^{\text{ext}},\quad \quad \quad \quad \ D= \epsilon E \quad \quad \text{in } Q^{\text{ext}} \label{4022}
\end{align}
for the magnetic permeability $\mu>0$ and the dielectric permittivity $\epsilon>0$ in the perfect conductor $\mathbb{R}^3 \setminus \overline{\Omega}$. Indeed, under this assumption the trivial relations \eqref{exteriorfinal} and the equation \eqref{ewoejfsejf} immediately imply that also the quantities $H$, $D$ and $\rho_c$ vanish in $Q^{\text{ext}}$, cf.\ for example \cite[Chapter 1: Part A: §4.2.4.3]{dautraylions}. In this case the whole electromagnetic subsystem in the exterior domain becomes trivial and it is sufficient to examine the problem in the interior domain. Thus, in the system \eqref{2919}--\eqref{2908}, in which the condition \eqref{4022} is assumed implicitly, all electromagnetic equations in the exterior domain are neglected. However, we point out that there does not seem to be a physical argument for the validity of the linear relations \eqref{4022} in a perfect conductor. For this reason we here decided to formulate the system \eqref{w3fejwejojjo}--\eqref{bic5} in a more general form, without the condition \eqref{4022} and with the electromagnetic subsystem in the exterior domain.

Furthermore, Gauss's law \eqref{ewoejfsejf} in the fluid domain is not included in the system \eqref{2919}--\eqref{2908}. This is explained as follows: In mathematical works in magnetohydrodynamics the (reduced) Maxwell system is commonly compressed into two equations for the magnetic induction $B$, Gauss's law for magnetism \eqref{jweo0jv0wsjd0} and the so-called induction equation,
\begin{align}
\operatorname{div} B = 0 \quad \text{in } Q^f, \quad \quad \partial _{t} B + \nabla \times (B \times u) + \frac{1}{\sigma} \nabla \times \left( \frac{1}{\mu} \nabla \times \left( B \right) -J \right) = 0 \quad \text{in } Q^f. \label{307}
\end{align}
Indeed, the latter equation results directly from a combination of Ohm's law \eqref{524}, Ampère's law \eqref{w3fejwejojjo} and the Maxwell-Faraday equation \eqref{ewofjweojf}. The idea behind this further reduction is that the unknown $B$ may be determined independently of all the other unknowns from the Maxwell system in the fluid domain. After determining $B$, we have the magnetic field $H$ given explicitly from the relation \eqref{523} and consequently also the electric current density $j$ from Ampère's law \eqref{w3fejwejojjo}. Subsequently, the electric field $E$ and the electric induction $D$ can be computed directly from Ohm's law \eqref{524} and the relation \eqref{523} and we can use Gauss's law \eqref{ewoejfsejf} to immediately obtain the density of electric charges $\rho _c$. Therefore, the Maxwell system in the fluid domain can be solved by solving only the system \eqref{307} and in particular Gauss's law \eqref{ewoejfsejf} becomes superfluous for the mathematical analysis. For this reason Gauss's law in the fluid domain is neglected in the system \eqref{2919}--\eqref{2908}.

As opposed to in the fluid domain, Gauss's law \eqref{ewoejfsejf} cannot be dropped in the solid domain. Indeed, in the solid domain $\rho_c$ is equal to $0$ (cf. \eqref{4031}), while $E$ cannot be determined via Ohm's law, so Gauss's law rather constitutes a condition required for determining $D$ and $E$ than for determining $\rho_c$. We remark that in the system \eqref{2919}--\eqref{2908} this equation is expressed in terms of $E$ instead of $D$, cf.\ \eqref{2918}. Thus the only remaining equation involving $D$ is the linear relation between $D$ and $E$ in \eqref{523}, which therefore becomes redundant for the analysis and consequently does not appear in the system \eqref{2919}--\eqref{2908}.

The linear relation between $B$ and $H$ in \eqref{523} instead constitutes a crucial component of the system \eqref{2919}--\eqref{2908}. In the latter system, however, it is modified in the sense that $\mu$ is assumed to take the same value in both $Q^f$ and $Q^s$, i.e.\ $\mu$ is a constant in the whole domain $Q$, cf.\ \eqref{2911}. This modification, which is only justified if the magnetic permeability in $Q^f$ and $Q^s$ is (almost) the same, is made for purely mathematical reasons. Namely, in combination with the interface conditions for $H$ and $B$ on $\partial S(t)$ in \eqref{bic2} and \eqref{bic3}, it ensures continuity of $B$ across the fluid-solid interface. This is necessary for the weak formulation of the system in \cite[Definition 2.1]{compressiblepaper} below, in which the magnetic induction is assumed to be a Sobolev function over the whole domain $Q$.

Moreover, we remark that the interface conditions for $H$ on $\partial \Omega$ in \eqref{bic2} as well as the interface conditions \eqref{bic4} on $D$ do not appear in the system \eqref{2919}--\eqref{2908} as they do not enter its weak formulation in \cite[Definition 2.1]{compressiblepaper}. Finally, we point out that also the conditions \eqref{4023} on $J$ are left out of the system \eqref{2919}--\eqref{2908} as the analysis in \cite{compressiblepaper} can also be carried out without them. This concludes the derivation of the system \eqref{2919}--\eqref{2908}.

\section{Existence of weak solutions} \label{exresult}

The remainder of the article is dedicated to a brief summary of the existence result on weak solutions to the system \eqref{2919}--\eqref{2908} which was published by the first author in \cite{compressiblepaper}, see also \cite{thesis}. 
We begin by introducing some additional notation required for the definition of weak solutions. The rigidity of the solid bodies implies the existence of orientation preserving isometries $\eta^i(t):\mathbb{R}^3 \rightarrow \mathbb{R}^3$ such that the position $S^i(t)$ of the $i$-th body at the time $t \in [0,T]$ may be expressed as
\begin{align}
S^i(t) := \eta^i\left(t, S_0^i \right), \nonumber
\end{align}

where $S_0^i$ denotes the corresponding initial position. The solid region $S(t)$ at time $t \in [0,T]$ can then be expressed with the help of the set-valued function
\begin{align}
S:\ [0,T] \rightarrow 2^{\mathbb{R}^3},\quad \quad S(t) := \eta\left(t, S_0 \right), \nonumber
\end{align}

where $S_0 := \bigcup_{i=1}^N S_0^i$ and
\begin{align}
\eta(t,\cdot):\ S_0 \rightarrow \mathbb{R}^3,\quad \quad \eta(t,\cdot)|_{S^i_0} := \eta^i(t,\cdot) . \nonumber
\end{align}

In order to establish a suitable connection between the isometries $\eta^i(t)$ and the velocity field $u$ in the problem, we demand compatibility of $u$ with the system $\{S_0^i, \eta^i\}_{i=1}^N$, meaning that each $\eta^i$, $i=1,...,N$, constitutes the Carathéodory solution to an initial value problem
\begin{align}
\frac{d\eta^i(t,x)}{dt} = u^{s^i}\left( t, \eta^i(t,x) \right), \quad \quad \eta^i(0,x) = x \label{-345}
\end{align}

for some rigid velocity field $u^{s^i}(t,\cdot)$ satisfying
\begin{align}
u(t,\cdot) = u^{s^i}(t,\cdot) \quad \quad \text{a.e. in } S^i(t)\ \text{for a.a. } t \in [0,T]. \label{-344}
\end{align}

The compatibility between $u$ and $\{S_0^i, \eta^i\}_{i=1}^N$ guarantees that the rigid bodies travel along the flow curves of $u$ and cannot penetrate each other, cf.\ \cite[Lemma 3.1, Lemma 3.2]{feireisl}. Finally, by
\begin{align}
H_\text{div}^{1}\left(\Omega \right) := \left\lbrace b \in H^1(\Omega):\ \operatorname{div} b = 0\ \text{in } \Omega \right\rbrace,\quad \quad L_\text{div}^{2}\left(\Omega \right) := \left\lbrace b \in L^2(\Omega):\ \operatorname{div} b = 0\ \text{in } \mathcal{D}'\left(\Omega \right)\right\rbrace \nonumber
\end{align}

we denote the space of $H^1(\Omega)$-functions and $L^2(\Omega)$-functions, respectively, which are additionally divergence-free and, for the weak formulation of the momentum equation and the induction equation, we introduce the test function spaces
\begin{align}
\mathcal{Z}(S) :=& \left\lbrace \phi \in \mathcal{D}\left([0,T)\times \Omega \right):\ \mathbb{D}(\phi) = 0\quad \text{in an open neighborhood of } \overline{Q^s(S)} \right\rbrace, \label{-339} \\
\mathcal{Y}(S) :=& \left\lbrace b \in \mathcal{D}\left([0,T)\times \Omega \right):\ \operatorname{curl}b = 0\quad \text{in an open neighborhood of } \overline{Q^s(S)} \right\rbrace. \label{-340}
\end{align}

Our definition of weak solutions to the system \eqref{2919}--\eqref{2908} reads as follows.
\begin{definition} \cite[Definition 2.1]{compressiblepaper}
\label{weaksolutionscompressible}
Let $T > 0$, let $\Omega \subset \mathbb{R}^3$ be a bounded domain and let $S_0 = \bigcup_{i=1}^N S_0^i$, where $S_0^i \subset \Omega$ for $i=1,...,N \in \mathbb{N}$ are bounded domains such that
\begin{align}
\emptyset \neq S_0^i \ \text{is open and connected, } |\partial S_0^i| = 0 \text{ and } \overline{S_0^i} \bigcap \overline{S_0^j} = \emptyset \quad \quad \forall i,j=1,...,N,\ i \neq j. \label{-326}
\end{align}
Assume $\nu,\lambda, a, \gamma,\sigma,\mu \in \mathbb{R}$ to satisfy
\begin{align}
\nu, a, \sigma,\mu > 0,\quad \nu + \lambda \geq 0, \quad \gamma > \frac{3}{2}, \label{-324}
\end{align}
consider some external data $g,J \in L^\infty (Q)$ and consider some initial data $0 \leq \rho_0 \in L^\gamma(\Omega)$, $(\rho u)_0 \in L^1(\Omega)$ and $B_0 \in L^2_{\operatorname{div}}(\Omega)$ satisfying
\begin{align}
\frac{\left|(\rho u)_0\right|^2}{\rho_0} \in L^1(\Omega),\quad (\rho u)_0 = 0 \ \ \text{a.e. in } \left\lbrace x \in \Omega:\ \rho_0(x) = 0 \right\rbrace,\quad B_0 \cdot \operatorname{n} = 0\ \text{on } \partial \Omega. \label{-325}
\end{align}
Then the system \eqref{2919}--\eqref{2908} is said to admit a weak solution on $[0,T]$ if there exists a function
\begin{align}
\eta:\ [0,T] \times S_0 \rightarrow \mathbb{R}^3,\quad \left. \eta(t,\cdot) \right|_{S_0^i} = \eta^i(t,\cdot) \quad \quad \forall i=1,...,N,\ t\in [0,T], \label{-159}
\end{align}
where each $\eta^i(t,\cdot): \mathbb{R}^3 \rightarrow \mathbb{R}^3$ denotes an orientation preserving isometry, and if there exist functions
\begin{align}
0 \leq \rho &\in L^\infty \left( 0,T;L^\gamma (\Omega;\mathbb{R}) \right) \bigcap C\left([0,T];L^{1}(\Omega;\mathbb{R}) \right), \label{-211} \\
u &\in \left\lbrace \phi \in L^2\left(0,T;H_0^{1}\left(\Omega;\mathbb{R}^3 \right)\right):\ \mathbb{D}(\phi) = 0 \ \text{in } Q^s(S) \right\rbrace, \label{-213} \\
B &\in \left\lbrace b \in L^\infty \left(0,T;L^2\left(\Omega; \mathbb{R}^3\right)\right) \bigcap L^2\left(0,T;H_{\operatorname{div}}^{1}\left(\Omega;\mathbb{R}^3\right)\right):\ \operatorname{curl}b = 0 \ \text{in } Q^s(S),\ b \cdot \operatorname{n} = 0 \ \text{on } \partial \Omega \right\rbrace, \label{-212}
\end{align}
where $S = S(\cdot) = \eta(\cdot, S_0)$, such that $\rho$ and $u$, extended by $0$ in $\mathbb{R}^3 \setminus \Omega$, satisfy the continuity equation 
\begin{align}
-\int_0^{T} \int _{\Omega} \rho \partial_t \psi dxdt - \int_{\Omega} \rho_0 \psi(0,x )\ dx =& \int_0^{T}\int_\Omega (\rho u) \cdot \nabla \psi \ dxdt \label{-214}
\end{align}
for all $\psi \in \mathcal{D}([0,T)\times \Omega)$ as well as the renormalized continuity equation
\begin{align}
\partial_t \zeta (\rho) + \operatorname{div} \left( \zeta \left(\rho \right)u \right) + \left[ \zeta'\left(\rho \right)\rho - \zeta \left(\rho \right) \right] \operatorname{div} u = 0 \quad &\text{in } \mathcal{D}' \left((0,T) \times \mathbb{R}^3 \right), \label{-215}
\end{align}
for any
\begin{align}
\zeta \in C^1\left( [0,\infty) \right):\quad \left|\zeta'(r)\right| \leq cr^{\lambda _1} \quad \forall r \geq 1, \quad \quad \text{where } c>0,\ \lambda_1 > -1, \label{-216}
\end{align}
such that the momentum equation and the induction equation,
\begin{align}
- \int_0^T \int_\Omega \rho u \cdot \partial_t \phi \ dxdt - \int_\Omega (\rho u)_0 \cdot \phi (0,x)\ dx =& \int _0^T \int _\Omega \left( \rho u \otimes u \right): \mathbb{D} (\phi) + a\rho^\gamma \operatorname{div} \phi - 2\nu \mathbb{D}(u):\mathbb{D} (\phi) \nonumber \\
&- \lambda \operatorname{div} u \operatorname{div} \phi + \rho g\cdot \phi + \frac{1}{\mu}\left( \operatorname{curl}B \times B \right) \cdot \phi \ dxdt, \label{-217} \\
- \int _0^T \int_\Omega B \cdot \partial_t b\ dxdt - \int_\Omega B_0 \cdot b(0,x)\ dx =& \int _0^T \int _\Omega \left[ -\frac{1}{\sigma \mu} \operatorname{curl} B + u \times B + \frac{1}{\sigma} J \right] \cdot \operatorname{curl} b \ dxdt, \label{-218}
\end{align}
are satisfied for any $\phi \in \mathcal{Z}(S)$ and any $b\in \mathcal{Y}(S)$ and, finally, such that the system $\lbrace S_0^i, \eta^i \rbrace_{i=1}^N$ is compatible with the velocity field $u$.
\end{definition}

The existence of weak solutions to the system \eqref{2919}--\eqref{2908} in the sense of Definition \ref{weaksolutionscompressible} is guaranteed by the following result.
\begin{theorem} \cite[Theorem 2.3]{compressiblepaper}
\label{mainresultcompressible}
Let $T > 0$, assume $\Omega \subset \mathbb{R}^3$ to be a simply connected domain of class $C^{2,\xi}$, $\xi \in (0,1)$, and assume $S_0^i \subset \Omega$, $i=1,...,N \in \mathbb{N}$ to be bounded domains of class $C^2$ which satisfy the conditions \eqref{-326}. Assume moreover the coefficients $\sigma, \mu, \nu, \lambda, a, \gamma \in \mathbb{R}$ to satisfy the conditions \eqref{-324} and the data $g,J \in L^\infty(Q)$, $\rho_0 \in L^\gamma (\Omega)$, $(\rho u)_0 \in L^1(\Omega)$ and $B_0 \in L^2_{\operatorname{div}}(\Omega)$ to satisfy the conditions \eqref{-325}. Then the system \eqref{2919}--\eqref{2908} admits a weak solution $(\eta,\rho,u,B)$ on $[0,T]$ in the sense of Definition \ref{weaksolutionscompressible} which in addition satisfies the energy inequality
\begin{align}
&\int_\Omega \frac{1}{2} \rho (\tau)\left|u (\tau)\right|^2 + \frac{a}{\gamma - 1} \rho ^\gamma(t) + \frac{1}{2\mu} \left| B(\tau) \right|^2\ dx + \int_0^\tau \int_\Omega 2\nu |\mathbb{D} \left(u \right)|^2 + \lambda \left| \operatorname{div}u \right|^2 + \frac{1}{\sigma \mu^2} \left| \operatorname{curl} B \right|^2\ dxdt \nonumber \\
\leq& \int_\Omega  \frac{1}{2} \frac{\left|\left(\rho u \right)_{0}\right|^2}{\rho_{0}} + \frac{a}{\gamma - 1} \rho_{0}^\gamma + \frac{1}{2\mu} \left| B_{0} \right|^2\ dx +\int_{0}^\tau\int_\Omega \rho g \cdot u + \frac{1}{\sigma \mu} J \cdot \operatorname{curl} B \ dxdt \label{-220}
\end{align}
for almost all $\tau \in [0,T]$.
\end{theorem}

\textbf{Proof}

As a courtesy to the readers we sketch the main ideas of the (quite technical) proof of Theorem \ref{mainresultcompressible}; the whole proof is given in \cite{compressiblepaper} and - with some additional details - in {\cite{thesis}. The proof is accomplished via an approximation method: First, an easily solvable approximation to the original system is constructed. After solving the approximate system, a solution to the original problem is then recovered by passing to the limit in the approximation. Our approximation consists of five different levels, characterized by five parameters $\Delta t>0$, $n,m \in \mathbb{N}$, $\epsilon, \alpha >0$:
\begin{itemize}
\item On the $\Delta t$-level, the induction equation is discretized with respect to the time variable via the Rothe method (see \cite[Section 8.2]{roubicek}), while the mechanical part of the problem is split up into a series of time-dependent problems on the small intervals between the discrete times. To this end we fix $\Delta t>0$ with $\frac{T}{\Delta t}\in \mathbb{N}$ and split up the interval $[0,T]$ into the discrete times $k\Delta t$, $k=1,...,\frac{T}{\Delta t}$.
\item On the $n$-level, a Galerkin method is carried out in order to solve the approximate momentum equation. 
\item The approximation levels associated to $m, \epsilon$ and $\alpha$ correspond to the approximation used in \cite{feireisl} for the purely mechanical problem: The $m$-level describes a penalization method which allows us to pass from a fluid-only system to a system containing both a fluid and rigid bodies. On the $\epsilon$- and $\alpha$-levels, the system is regularized through the addition of multiple regularization terms as well as an artificial pressure. 
\end{itemize}
The demand for a time discretization on the $\Delta t$-level can be explained in the following way: The fact that the test functions in both the momentum equation and the induction equation depend on the solution of the system (cf.\ \eqref{-339}, \eqref{-340}) prevents us from solving all of the equations in the original system simultaneously. In the case of the momentum equation this problem is circumvented via the application of a penalization method (see the $m$-level below). In the case of the induction equation, however, no similar method appears to be available. Instead we decouple the system via a time discretization, which allows us to solve the equations one after another by using time-lagging functions in the coupling terms. In this way, at each discrete time $k\Delta t$, $k=1,...,\frac{T}{\Delta t}$, in our discretization we are able to first determine the position $S_{\Delta t, k}(k \Delta t)$ of the solid and subsequently solve the discrete induction equation at this specific time with test functions chosen accordingly.} More specifically, as discrete approximations of the test functions \eqref{-340} we consider test functions $b \in H^1(\Omega)$ satisfying
\begin{align}
\text{curl}\ b = 0\ \text{in } S_{\Delta t, k}(k\Delta t) \bigcap \Omega ,\quad b \cdot \text{n}|_{\partial \Omega} = 0. \nonumber
\end{align}

Then, a discretized version of the induction equation of the form
\begin{align}
-\int_\Omega \frac{B_{\Delta t}^k - B_{\Delta t}^{k-1}}{\Delta t} \cdot b\ dx =& \int_\Omega \left[ \frac{1}{\sigma \mu} \operatorname{curl} B_{\Delta t}^k - u_{\Delta t}^k \times B_{\Delta t}^{k-1} - \frac{1}{\sigma} J_{\Delta t}^k \right] \cdot \operatorname{curl} b\ dx, \label{116}
\end{align}
wherein $B_{\Delta t}^k$, $u_{\Delta t}^k$ and $J_{\Delta t}^k$ denote discrete approximations of the unknown $B$ and $u$ and the given force $J$, can be solved under the application of classical methods. The deployment of a time discretization, however, leads to further problems. More precisely, the well-known techniques for showing non-negativity of the density in the construction of weak solutions to the compressible Navier-Stokes equations (see \cite[Sections 7.6.5, 7.6.6, 7.7.4.2]{novotnystraskraba}) do not seem to be compatible with discrete equations. In particular, it does not seem possible to discretize the mechanical part of our system in such a way that a meaningful energy inequality can be derived, which in turn is required for obtaining uniform bounds of the approximate solution and passing to the limit in the discretization.

This issue motivates the use of a hybrid approximation scheme, in which the induction equation is discretized as in \eqref{116} while the mechanical equations are treated as continuous equations on the small intervals $[(k-1)\Delta t,k\Delta t]$ between the consecutive discrete time points. Consequently, the problem of the solution-dependence of the test functions in the induction equation can be handled as described above while the non-negativity of the density can be achieved under exploitation of the classical existence theory for the compressible Navier-Stokes equations. A suitable overall energy inequality for the approximate problem can then be derived under the consideration of piecewise linear interpolants of the discrete functions, which allows us to deduce a continuous energy estimate from the discrete induction equation. This can subsequently be added together with the corresponding mechanical estimate to obtain the full energy inequality.

The Galerkin method on the $n$-level constitutes a classical procedure in the solving of the continuous compressible Navier-Stokes equations. In addition, the higher regularity of the velocity field on the Galerkin level is of benefit during the limit passage in the time discretization.

The penalization method on the $m$-level is what allows us to handle the solution dependent test functions \eqref{-339} in the momentum equation. For the same purpose it has been used previously in the purely mechanical problem in, for example, \cite{feireisl} in the compressible case and \cite{tucsnak} in the incompressible setting. In this method, a fluid is considered in the whole domain and the rigid bodies are approximated by letting the viscosity of this fluid rise to infinity in the later solid region. The approximate all-fluid problem can then be solved with classical test functions independent of the solution and it suffices to switch to the solution-dependent test functions \eqref{-339} in the limit passage with respect to $m \rightarrow \infty$. For the mathematical implementation of this penalization method, we define the approximate solid region $S(t)$ at any time $t \in [0,T]$ through the flow curves of a suitable regularization $R_\delta [u]$, $\delta > 0$, of the (fluid) velocity $u$: Namely, we set
\begin{align}
S(t) := \bigcup_{i=1}^N S^i(t), \nonumber
\end{align}

where
\begin{align}
S^i(t) := \left( O^i (t) \right)^\delta = \left\lbrace x \in \mathbb{R}^3:\ \operatorname{dist}\left(x, O^i(t) \right) < \delta \right\rbrace,\quad O^i (t) := \eta\left(t,O^i\right), \nonumber
\end{align}

denotes the position of the approximate $i$-th solid body, $\eta$ is determined via the initial value problem
\begin{align}
\frac{d}{dt}\eta(t,x) =& R_\delta \left[ u \right] \left( t, \eta(t,x) \right), \quad \quad \eta (0,x) = x \label{flowcurves}
\end{align}

and 
\begin{align}
O^i := \left( S_0^i \right)_\delta := \left\lbrace x \in S_0^i:\ \operatorname{dist} \left(x, \partial S_0^i \right) > \delta \right\rbrace \nonumber 
\end{align}

denotes the $\delta$-kernel of the initial position $S_0^i$ of the $i$-th body. We point out that the regularized velocity field $R_\delta[u]$ can be chosen such that the implication
\begin{align}
\mathbb{D}(u(t,\cdot)) = 0\quad \text{in } U \quad \quad \Rightarrow \quad \quad R_\delta [u](t,\cdot) = u(t,\cdot) \quad \text{in } U_\delta = \left\lbrace x \in U:\ \operatorname{dist}\left( x, \partial U \right)>\delta \right\rbrace \label{-348}
\end{align}
holds true for any domain $U \subset \mathbb{R}^3$. We further denote by
\begin{align}
\chi (t,x) := \operatorname{dist} \left(x, \overline{ \mathbb{R}^3 \setminus S(t)} \right) - \operatorname{dist} \left( x, \overline{S(t)} \right) \nonumber
\end{align}
the signed distance function of $S(t)$ and fix some convex function $H \in C^\infty (\mathbb{R})$ such that $H(z) = 0$ for $z \in (-\infty, 0]$ and $H(z)> 0$ otherwise. Then, we replace the viscosity coefficients $\nu$ and $\lambda$ in the momentum equation \eqref{-217} by the quantities
\begin{align}
\nu \left( \chi \right) := \nu + m H\left(\chi \right),\quad \quad \lambda \left( \chi \right) := \lambda + m H\left(\chi \right), \nonumber
\end{align}

which are referred to as variable viscosity coefficients. In the limit passage with respect to $m \rightarrow \infty$, the coefficients $\nu(\chi)$ and $\lambda(\chi)$ rise to infinity in the (approximate) solid region. The associated energy inequality then shows that, in each solid body $S^i(t)$, the limit velocity $u$ has a vanishing symmetric gradient $\mathbb{D}(u)=0$ and is thus equal to a rigid velocity field $u^{S_i}$. Since, due to the relation \eqref{flowcurves}, the bodies $S^i(t)$ in the limit are again determined via the flow curves of $R_\delta[u]$, the implication \eqref{-348} shows that $R_\delta [u]=u$ in the sets $O^i(t) \subset S^i(t)$. It follows that
\begin{align}
u^{S_i} = u = R_\delta[u] \quad \quad \text{in } O^i(t). \nonumber
\end{align}
This implies rigidity of the bodies $S^i(t)$, the motion of which is determined by $R_\delta[u]$, as well as compatibility of $u$ with the system $\{S_0^i,\eta^i\}_{i=1}^N$ for certain orientation preserving isometries $\eta^i(t,\cdot):\mathbb{R}^3 \rightarrow \mathbb{R}^3$.

On the $\epsilon$-level, several regularization methods with different purposes are combined into one step. The continuity equation is regularized by the addition of the Laplacian $\epsilon \Delta \rho$ of the density to its right-hand side. This parabolic regularization constitutes the classical approach from the theory of the compressible Navier-Stokes equations for the construction of a non-negative density, see \cite[Sections 7.6.5, 7.6.6, 7.7.4.2]{novotnystraskraba}. In order to compensate for this additional term in the derivation of the energy inequality, the momentum equation is moreover supplemented by the quantity $\epsilon \nabla u \nabla \rho$. Furthermore, the quantity
\begin{align}
\epsilon |u|^2 u \label{regterm}
\end{align}

is added to the momentum equation. This is due to the fact that, while in the continuous system the mixed terms from the momentum equation and the induction equation cancel each other in the energy inequality, the same is not true in our hybrid approximation. Therefore, in order to obtain uniform bounds for the approximate solution, these terms need to be controlled by the non-negative left-hand side of the energy inequality. This is possible via an estimate of the form
\begin{align}
\int_\Omega \frac{1}{\mu}\left( u_{\Delta t}^{k-1} \times B_{\Delta t}^{k-1} \right) \cdot \operatorname{curl} B_{\Delta t}^{k} \ dx \leq \frac{1}{\sqrt{\mu} \epsilon} \left\| B_{\Delta t}^{k-1} \right\|_{L^2(\Omega)}^2 + \frac{\epsilon}{8} \left\| u_{\Delta t}^{k-1} \right\|_{L^4(\Omega)}^4 + \frac{\epsilon}{8\mu^3} \left\| \operatorname{curl}B_{\Delta t}^k \right\|_{L^4(\Omega)}^4 \label{-365}
\end{align}

for the mixed term in the discrete induction equation \eqref{116} and a similar estimate for the mixed term in the momentum equation. Here, the term $\frac{\epsilon}{8}\| u_{\Delta t}^{k-1} \|_{L^4(\Omega)}^4$ can be absorbed by the left-hand side of the energy inequality on the $\Delta t$-level thanks to the regularization term \eqref{regterm}. In order to control the term $\frac{1}{\epsilon 8\mu^3}\| \operatorname{curl} B_{\Delta t}^k \|^4_{L^4(\Omega)}$ in the estimate \eqref{-365} in the same way, the induction equation is regularized similarly through the addition of the $4$-double-curl
\begin{align}
\frac{\epsilon}{\mu^2} \text{curl} \left(| \text{curl} B |^2 \text{curl}B\right) \nonumber
\end{align}

of the magnetic induction. The term $\frac{1}{\mu \epsilon}\|B_{\Delta t}^{k-1}\|^2_{L^2(\Omega)}$ in the estimate \eqref{-365} can then simply be controlled via the Gronwall lemma, which leads to the desired bounds uniform with respect to $\Delta t$. The induction equation is further regularized through the addition of the term
\begin{align}
\epsilon \operatorname{curl} (\Delta (\operatorname{curl} B)). \nonumber
\end{align}

With the help of this term, the induction equation can be expressed via a weakly continuous and coercive operator, which guarantees its solvability.

Finally, the addition of an artificial pressure to the momentum equation on the $\alpha$-level is also well-known from the classical theory for the compressible Navier-Stokes system, see for example \cite[Section 7.3.8]{novotnystraskraba}. This artificial pressure, which is chosen of the form $\alpha \rho^\beta$ for a sufficiently large $\beta > \max\{4,\gamma\}$, helps us to pass to the limit in the quantity $\epsilon ( \nabla u \nabla \rho)$ from the $\epsilon$-level by guaranteeing a higher integrability of the density, cf.\ \cite[Section 7.8.2]{novotnystraskraba}. The higher integrability moreover enables the use of the regularization technique by DiPerna and Lions (cf.\ \cite[Lemma 6.8, Lemma 6.9]{novotnystraskraba}) and thus facilitates the limit passage with respect to $\epsilon \rightarrow 0$.

\section{Appendix}

Here we present some calculations needed for the derivation of the boundary and interface conditions in Section \ref{bicond}. These are mainly elementary and follow the ideas of the standard pill box argument in the physics literature. Here we present the calculations for the case of a curved instead of a flat interface.

\subsection{Proof of the estimate \eqref{wefujwe09uuj09}} \label{etangential}

In this section we prove the estimate \eqref{wefujwe09uuj09}, used in the derivation of the boundary and interface conditions \eqref{831mod} for the tangential component of $H$ in Section \ref{bicond}. To this end we first exploit the fact that $\phi$ is twice continuously differentiable with $\phi(0)=\phi'(0)=0$ and $H$ is continuously differentiable in both the fluid and the solid domain, cf.\ \eqref{mathassumption2}. This allows us to use the fundamental theorem of calculus and calculate the integrals on the left-hand side of the integrated version \eqref{maxwellfaradayint} of Ampère's law as
\begin{align}
\int_{S_1} H \cdot ds =& \int_{-\Delta l}^{\Delta l} H\left(\begin{pmatrix}\Delta s\\ \phi (\Delta s) + \zeta \\ 0\end{pmatrix} \right) \cdot \left(
\begin{array}{c}
0\\
1\\
0\\
\end{array}
\right)\ d\zeta \nonumber \\
=& \int_{-\Delta l}^{\Delta l} H\left(\begin{pmatrix}0\\ \zeta \\ 0\end{pmatrix}\right)\cdot \left(
\begin{array}{c}
0\\
1\\
0\\
\end{array}
\right) + \int_0^{\Delta s} \frac{d}{d\xi} H\left(\begin{pmatrix} \xi \\ \phi(\xi) + \zeta \\0\end{pmatrix} \right)\cdot \left(
\begin{array}{c}
0\\
1\\
0\\
\end{array}
\right)\ d\xi d\zeta, \label{S1} \\
\int_{S_2} H \cdot ds =& \int_{-\Delta s}^{\Delta s} H\left( \begin{pmatrix} -\theta \\ \phi(-\theta) + \Delta l \\ 0\end{pmatrix}\right) \cdot \left(
\begin{array}{c}
-1\\
-\phi'(-\theta)\\
0\\
\end{array}
\right)\ d\theta \nonumber \\
=& \int_{-\Delta s}^{\Delta s} H\left(\begin{pmatrix} 0\\ \Delta l\\ 0\end{pmatrix}\right) \cdot \left(
\begin{array}{c}
-1\\
0\\
0\\
\end{array}
\right) + \int_0^\theta \frac{d}{d\xi} \left[ H\left( \begin{pmatrix} -\xi \\ \phi(-\xi)+\Delta l \\0 \end{pmatrix}\right)\cdot \left(
\begin{array}{c}
-1\\
-\phi'(-\xi)\\
0\\
\end{array}
\right) \right]\ d\xi d\theta, \label{S2} \\
\int_{S_3} H \cdot ds =& \int_{-\Delta l}^{\Delta l}H\left(\begin{pmatrix} 0 \\ - \zeta \\ 0\end{pmatrix}\right)\cdot \left(
\begin{array}{c}
0\\
-1\\
0\\
\end{array}
\right) + \int_0^{\Delta s} \frac{d}{d\xi} H\left(\begin{pmatrix}-\xi \\ \phi(-\xi) - \zeta \\0\end{pmatrix}\right)\cdot \left(
\begin{array}{c}
0\\
-1\\
0\\
\end{array}
\right)\ d\xi d\zeta, \label{S3} \\
\int_{S_4} H \cdot ds =& \int_{-\Delta s}^{\Delta s} H\left(\begin{pmatrix} 0\\ - \Delta l\\ 0\end{pmatrix}\right) \cdot \left(
\begin{array}{c}
1\\
0\\
0\\
\end{array}
\right) + \int_0^\theta \frac{d}{d\xi} \left[ H\left( \begin{pmatrix} \xi \\ \phi(\xi)-\Delta l \\0 \end{pmatrix}\right)\cdot \left(
\begin{array}{c}
1\\
\phi'(\xi)\\
0\\
\end{array}
\right) \right]\ d\xi d\theta. \label{S4}
\end{align}
We remark that for the first terms on the right-hand sides of the equations \eqref{S1} and \eqref{S3}, respectively, it holds that
\begin{align}
\int_{-\Delta l}^{\Delta l} H\left(\begin{pmatrix}0\\ \zeta \\ 0\end{pmatrix}\right)\cdot \left(
\begin{array}{c}
0\\
1\\
0\\
\end{array}
\right)\ d\zeta + \int_{-\Delta l}^{\Delta l}H\left(\begin{pmatrix} 0 \\ - \zeta \\ 0\end{pmatrix}\right)\cdot \left(
\begin{array}{c}
0\\
-1\\
0\\
\end{array}
\right)\ d\zeta = 0. \label{0}
\end{align}
With the identities \eqref{S1}--\eqref{0} at hand we rearrange the equation \eqref{maxwellfaradayint}. From the boundedness of $H$, $D$, $J$ and $\phi$ as well as their derivatives (cf.\ \eqref{mathassumption3}) we then infer that
\begin{align}
&\left| \int_{-\Delta s}^{\Delta s} H\left(\begin{pmatrix}0\\ \Delta l \\ 0\end{pmatrix}\right)\cdot \left(
\begin{array}{c}
-1\\
0\\
0\\
\end{array}
\right)\ d\theta + \int_{-\Delta s}^{\Delta s}H\left(\begin{pmatrix} 0 \\ - \Delta l \\ 0\end{pmatrix}\right)\cdot \left(
\begin{array}{c}
1\\
0\\
0\\
\end{array}
\right)\ d\theta - \int_{\Delta F} j \cdot \text{n}_x'\ dA \right| \nonumber \\
\leq& c \left( \Delta s \Delta l + (\Delta s)^2 \right) \nonumber
\end{align}
for some constant $c>0$ independent of $\Delta s$ and $\Delta l$. Dividing this inequality by $2\Delta s$ we conclude that
\begin{align}
\left| \tau_x \cdot H\left(\begin{pmatrix} 0\\ \Delta l \\0\end{pmatrix} \right) - \tau_x \cdot H\left(\begin{pmatrix}0\\ - \Delta l\\ 0 \end{pmatrix}\right) - \frac{1}{2\Delta s} K(\Delta F) \right| \leq c(\Delta l + \Delta s), \nonumber
\end{align}
where $\tau_x = \text{n}_x' \times \text{n}_x = (-1,0,0)^T$ and $K(\Delta F)$ is defined by \eqref{Kdef}, i.e.\ the desired inequality \eqref{wefujwe09uuj09}.

\subsection{Proof of the estimate \eqref{boundaryestimate}} \label{bnormal}

In this section we show the estimate \eqref{boundaryestimate}, needed in the deduction of the boundary and interface conditions \eqref{834mod} for the normal component of $D$ in Section \ref{bicond}. To this end we rewrite the integrals in the integrated version \eqref{rfrwweügiii} of Gauss's law under usage of the parametrizations \eqref{f43j0j4f04f}--\eqref{dr34f304f0i3}. We start by calculating, from the parametrization \eqref{f43j0j4f04f},
\begin{align}
\left| \frac{\partial \Gamma_{\text{top}}(s, \alpha)}{\partial s} \times \frac{\partial \Gamma_{\text{top}}(s, \alpha)}{\partial \alpha} \right| \text{n}_{\text{top}}(s, \alpha) = s \begin{pmatrix} \partial_1 \Phi (s \cos (\alpha), s \sin (\alpha)) \\ -1 \\ \partial _2\Phi (s \cos (\alpha), s \sin (\alpha)) \end{pmatrix}. \nonumber
\end{align}
This in combination with the parametrization (\ref{f43j0j4f04f}), the fundamental theorem of calculus - which may be applied since $D$ is continuously differentiable in both the fluid and the solid domain and $\Phi$ is twice continuously differentiable - and the identities $\Phi(0,0)=\partial_1 \Phi(0,0) = \partial_2 \Phi(0,0)=0$ allows us to calculate the first integral on the left-hand side of the identity \eqref{rfrwweügiii} as
\begin{align}
\int_{\Delta _\text{top}} D \cdot \text{n}_{\text{top}}\ dA =& \int_0^{2\pi} \int_0^r s D \left(\begin{pmatrix} s \cos(\alpha) \\ \Phi (s \cos (\alpha), s \sin (\alpha)) + \Delta l \\ s \sin(\alpha) \end{pmatrix}\right) \cdot \begin{pmatrix} \partial_1 \Phi (s \cos (\alpha), s \sin (\alpha)) \\ -1 \\ \partial _2\Phi (s \cos (\alpha), s \sin (\alpha)) \end{pmatrix}\ ds d\alpha \nonumber \\
=& \pi r^2 D \left( \begin{pmatrix} 0 \\ \Delta l \\ 0 \end{pmatrix} \right) \cdot \begin{pmatrix} 0 \\ -1 \\ 0 \end{pmatrix} \nonumber \\
+ \int_0^{2\pi} \int_0^r &s \int_0^s \frac{d}{d\xi} \left[ D \left( \begin{pmatrix} \xi \cos(\alpha) \\ \Phi (\xi \cos (\alpha), \xi \sin (\alpha)) + \Delta l \\ \xi \sin(\alpha) \end{pmatrix} \right) \cdot \begin{pmatrix} \partial_1 \Phi (\xi \cos (\alpha), \xi \sin (\alpha)) \\ -1 \\ \partial _2\Phi (\xi \cos (\alpha), \xi \sin (\alpha)) \end{pmatrix} \right]\ d\xi ds d\alpha. \label{ntop}
\end{align}
Analogously, for the second integral on the left-hand side of the identity \eqref{rfrwweügiii}, we calculate, from the parametrization (\ref{d3f3j0j4fjj}),
\begin{align}
&\int_{\Delta _\text{bottom}} D \cdot \text{n}_{\text{bottom}}\ dA = \pi r^2 D \left(\begin{pmatrix} 0 \\ -\Delta l \\ 0 \end{pmatrix}\right) \cdot \begin{pmatrix} 0 \\ 1 \\ 0 \end{pmatrix} \nonumber \\
& + \int_0^{2\pi} \int_0^r s \int_0^s \frac{d}{d\xi} \left[ D \left(\begin{pmatrix} \xi \cos(-\alpha) \\ \Phi (\xi \cos (-\alpha), \xi \sin (-\alpha)) - \Delta l \\ \xi \sin(-\alpha) \end{pmatrix}\right) \cdot \begin{pmatrix} -\partial_1 \Phi (\xi \cos (-\alpha), \xi \sin (-\alpha)) \\ 1 \\ -\partial _2\Phi (\xi \cos (-\alpha), \xi \sin (-\alpha)) \end{pmatrix} \right]\ d\xi ds d\alpha. \label{nbottom}
\end{align}
Finally, a similar calculation for the third integral on the left-hand side of \eqref{rfrwweügiii} under exploitation of the parametrization \eqref{dr34f304f0i3} leads to
\begin{align}
&\int_{\Delta _\text{side}} D \cdot \text{n}_{\text{side}}\ dA \nonumber \\
=& \int_{-\Delta l}^{\Delta l} \int_0^{2\pi} r \left[ D\left(\begin{pmatrix} 0 \\ h \\ 0 \end{pmatrix}\right) \cdot \begin{pmatrix}
-\cos (\alpha ) \\ 0 \\ -\sin(\alpha) \end{pmatrix} \right. \nonumber \\
&\quad \quad \quad \quad \quad \left.+ \int_0^r \frac{d}{d\xi} D \left( \begin{pmatrix} \xi \cos(\alpha) \\ \Phi (\xi \cos (\alpha), \xi \sin (\alpha)) + h \\ \xi \sin(\alpha) \end{pmatrix} \right) \cdot \begin{pmatrix}
-\cos (\alpha ) \\ 0 \\ -\sin(\alpha) \end{pmatrix} d\xi \right]\ d\alpha dh \nonumber \\
=& \int_{-\Delta l}^{\Delta l} \int_0^{2\pi} r \int_0^r \frac{d}{d\xi} D \left( \begin{pmatrix} \xi \cos(\alpha) \\ \Phi (\xi \cos (\alpha), \xi \sin (\alpha)) + h \\ \xi \sin(\alpha) \end{pmatrix} \right) \cdot \begin{pmatrix}
-\cos (\alpha) \\ 0 \\ -\sin(\alpha) \end{pmatrix}\ d\xi d\alpha dh. \label{nside}
\end{align}
We use the identities \eqref{ntop}, \eqref{nbottom} and \eqref{nside} to reformulate the terms on the left-hand side of the equation (\ref{rfrwweügiii}). Rearranging the resulting equation and using the fact that $D$ and $\phi$ as well as their derivatives are bounded we estimate
\begin{align}
\left| \pi r^2 D \left( \begin{pmatrix} 0 \\ \Delta l \\ 0 \end{pmatrix} \right) \cdot \begin{pmatrix} 0 \\ 1 \\ 0 \end{pmatrix} - \pi r^2 D \left(\begin{pmatrix} 0 \\ -\Delta l \\ 0 \end{pmatrix}\right) \cdot \begin{pmatrix} 0 \\ 1 \\ 0 \end{pmatrix} - \int_C \rho_c \ dx \right| \leq c \left(r^3 + r^2 \Delta l \right) \nonumber
\end{align}
for a constant $c>0$ independent of $r$ and $\Delta l$. Dividing this inequality by $\pi r^2$ we arrive at the estimate
\begin{align}
\left| \text{n}_x \cdot \left[ D\left( \begin{pmatrix} 0 \\ -\Delta l \\ 0 \end{pmatrix}\right) - D\left( \begin{pmatrix} 0 \\ \Delta l \\ 0 \end{pmatrix}\right) \right] - \frac{1}{\pi r^2}W(C) \right| \leq c (r + \Delta l), \nonumber
\end{align}
where $\text{n}_x = (0,1,0)^T$ on $\partial S(t)$ constitutes the outer unit normal vector in the origin $x=(0,0,0)^T$ of the local coordinate system and $W(C)$ is defined by \eqref{Wdef}. This proves the desired estimate \eqref{boundaryestimate}. 

\vspace{1cm}

\textbf{Acknowledgment}
The main essence of this work was obtained at the University of Würzburg and forms parts of the first author's PhD thesis \cite{thesis}. Some of the writing process was conducted while J.S. was working at the University of Prague and the Czech Academy of Sciences, respectively. J.S.\ acknowledges the support by Praemium Academiæ of \v{S}árka Ne\v{c}asová. The Institute of Mathematics, CAS is supported by RVO:67985840.

\end{document}